\documentclass[12pt]{article} 

\usepackage{epsfig} 
\usepackage{amsbsy}  
\usepackage{wasysym}
 
\newlength{\largfig}
\def\nul{\nu_\ell}

\def\ds#1{#1\kern-1ex\hbox{/}} 
\def\sl#1{#1\kern-1ex\hbox{/}} 
\def\dsh{h\kern-1.2ex /}

\def\beq{\begin{equation}} 
\def\eeq{\end{equation}} 
\def\eq{\beq\eeq} 
\def\beqn{\begin{eqnarray}} 
\def\eeqn{\end{eqnarray}} 
 
\def\lq{\left[} 
\def\rq{\right]}

\def\({\left(} 
\def\){\right)} 
 
\def\ba{\begin{eqnarray}} 
\def\ea{\end{eqnarray}} 
\def\bq{\begin{equation}} 
\def\eq{\end{equation}} 
 
\def\gsim{\mathrel{\raisebox{-.6ex}{$\stackrel{\textstyle>}{\sim}$}}}

\def\MB{{\cal M}_B}

\def\I{{\cal I}} 
\def\I{{\boldsymbol{I}}}

\def\asb{{}\ifmmode \bar{\alpha}_s \else $\bar{\alpha}_s$\fi} 
\def \as   {\ifmmode \alpha_s \else $\alpha_s$ \fi}

\def\al{\alpha}

\def\so3#1{\,{\rm S}_{1,\,3}\left(#1 \right)} 
\def\st2#1{\,{\rm S}_{2,\,2}\left(#1 \right)} 

\def\Re{\mathop{\rm Re}} 
\def\MVV{\widetilde{\cal M}_{V_1 V_2,\tau}\(q_1,q_2\)} 
\def\MVVi{\widetilde{\cal M}_{V_1 V_2,\tau}^{(i)}\(q_1,q_2\)} 
\def\MVVV{\widetilde{\cal M}_{V_1 V_2 V_3,\tau}\(q_1,q_2,q_3\)} 
\def\MVVVi{\widetilde{\cal M}_{V_1 V_2 V_3,\tau}^{(i)}\(q_1,q_2,q_3\)} 
\def\fortran{{\tt fortran}}
\def\MAPLE{{\tt MAPLE}}
\def\MADGRAPH{{\tt MadGraph}}
\def\wwdecay{e^+\nu_e \,\mu^-\bar\nu_\mu}
\def\wwjj{W^+W^-jj}
\def\Kfac{$K$ factor}

\def\sla#1{\ifmmode%
\setbox0=\hbox{$#1$}%
\setbox1=\hbox to\wd0{\hss$/$\hss}\else%
\setbox0=\hbox{#1}%
\setbox1=\hbox to\wd0{\hss/\hss}\fi%
#1\hskip-\wd0\box1 }

\newskip\humongous \humongous=0pt plus 1000pt minus 1000pt

\newif\ifdtup

\catcode`@=12 
 
\catcode`\@=11 
 
\@addtoreset{equation}{section} 

\def\theequation{\thesection.\arabic{equation}} 
 
\def\@normalsize{\@setsize\normalsize{15pt}\xiipt\@xiipt 
\abovedisplayskip 14pt plus3pt minus3pt%
\belowdisplayskip \abovedisplayskip 
\abovedisplayshortskip \z@ plus3pt%
\belowdisplayshortskip 7pt plus3.5pt minus0pt} 
 
\def\small{\@setsize\small{13.6pt}\xipt\@xipt 
\abovedisplayskip 13pt plus3pt minus3pt%
\belowdisplayskip \abovedisplayskip 
\abovedisplayshortskip \z@ plus3pt%
\belowdisplayshortskip 7pt plus3.5pt minus0pt 
\def\@listi{\parsep 4.5pt plus 2pt minus 1pt 
     \itemsep \parsep 
     \topsep 9pt plus 3pt minus 3pt}} 
 
\@twosidetrue

\catcode`\@=11  
\def\section{\@startsection{section}{1}{\z@}{3.5ex plus 1ex minus 
   .2ex}{2.3ex plus .2ex}{\large\bf}}

\def\thesection{\arabic{section}} 
\def\thesubsection{\arabic{section}.\arabic{subsection}} 
\def\thesubsubsection{\arabic{section}.\arabic{subsection}.\arabic{subsubsection}} 
 
\def\appendix{\setcounter{section}{0} 
 \def\thesection{\Alph{section}} 
 \def\theequation{\Alph{section}.\arabic{equation}} 
\def\thesubsection{\Alph{section}.\arabic{subsection}} 
\def\thesubsubsection{\Alph{section}.\arabic{subsection}.\arabic{subsubsection}} 
 
\def\section{\@startsection{section}{1}{\z@}{3.5ex plus 1ex minus 
   .2ex}{2.3ex plus .2ex}{\large\bf}} 
}

\newcommand{\ccaption}[2]{ 
  \begin{center} 
    \parbox{0.85\textwidth}{ 
      \caption[#1]{\small\it {#2}}} 
  \end{center}    } 
 
\def \ep{\epsilon} 
 
\def \to   {\mbox{$\rightarrow$}}

\def\ord#1{{\cal O}\(#1\)} 
 
\newcount\minutes 
\newcount\scratch 
 
\def\timestamp{%
\scratch=\time 
\divide\scratch by 60 
\edef\hours{\the\scratch} 
\multiply\scratch by 60 
\minutes=\time 
\advance\minutes by -\scratch 
---$\,$\hours:\null 
\ifnum\minutes< 10 0\fi 
\the\minutes}

\begin{document} 
\begin{titlepage} 
\nopagebreak 
{\flushright{ 
        \begin{minipage}{5cm} 
         Bicocca-FT-06-4 \\
         KA--TP--03--2006  \\       
         SFB--CPP--06--10 \\
        {\tt hep-ph/0603177}\hfill \\ 
        \end{minipage}        } 
 
} 
\vfill 
\begin{center} 
{\LARGE \bf \sc 
 \baselineskip 0.9cm 
Next-to-leading order QCD corrections to $W^+W^-$ production via\\
vector-boson fusion 

} 
\vskip 0.5cm  
{\large   
Barbara J\"ager$^1$, Carlo Oleari$^2$ and Dieter Zeppenfeld$^1$ 
}   
\vskip .2cm  
{$^1$ {\it Institut f\"ur Theoretische Physik, 
        Universit\"at Karlsruhe, P.O.Box 6980, 76128 Karlsruhe, Germany}
}\\ 
{$^2$ {\it Dipartimento di Fisica "G. Occhialini", 
        Universit\`a di Milano-Bicocca, 
        20126 Milano, Italy}}\\   
 
 \vskip 
1.3cm     
\end{center} 
 
\nopagebreak 
\begin{abstract}
Vector-boson fusion processes constitute an important class of reactions
at hadron colliders, both for signals and backgrounds of new physics
in the electroweak interactions. 
We consider what is commonly referred to as $W^+W^-$ production via
vector-boson fusion (with subsequent leptonic decay of the $W$s), or,
more precisely, $\wwdecay$ + 2 jets production in proton-proton scattering, 
with all resonant and non-resonant Feynman diagrams and spin correlations of
the final-state leptons included, in the phase-space regions which are 
dominated by $t$-channel electroweak-boson exchange.
We compute the next-to-leading order QCD corrections to this process, 
at order $\alpha^6\alpha_s$.
The QCD corrections are modest, changing total cross sections by less than
10\%. Remaining scale uncertainties are below 2\%. A fully-flexible
next-to-leading order partonic Monte Carlo program allows to demonstrate
these features for cross sections within typical vector-boson-fusion
acceptance cuts. Modest corrections are also found for distributions.
\end{abstract} 
\vfill 
\vfill 

\end{titlepage} 
\newpage

\section{Introduction} 
Vector-boson fusion (VBF) processes form a particularly
interesting class of scattering events from which one hopes to gain
insight into the dynamics of electroweak symmetry breaking. The most
prominent example is Higgs boson production, that is the process
$qq\,\to\, qqH$, which can be viewed as quark scattering via $t$-channel
exchange of a weak boson, with the Higgs boson radiated off the $W$ or $Z$
propagator. Alternatively, one may view this process as two weak bosons fusing
to form the Higgs boson. Higgs boson production via VBF has been studied
intensively as a tool for Higgs boson discovery~\cite{ATLAS,CMS} and the
measurement of Higgs boson couplings~\cite{Zeppenfeld:2000td} in
$pp$ collisions at the CERN Large Hadron Collider (LHC). The two scattered
quarks in a VBF process are usually visible as forward jets and greatly help
to distinguish these $Hjj$ events from backgrounds.

An important background to Higgs searches at the LHC, in particular to
the search for $H\,\to\, W^+W^-$ decays in VBF production, is caused by
continuum $W^+W^-$ production in VBF. 
The $qq \,\to\, qq\,W^+W^-$ process forms an
irreducible background in Higgs searches which ranges between 15\% and
3.5\% of the Higgs signal, for Higgs boson masses between 115 and
160~GeV~\cite{wbfhtoww}. In fact, the 
 kinematic distributions of
the two tagging jets, the suppression of gluon radiation in the central
region (due to the $t$-channel color-singlet exchange nature of the VBF
process) and many features of the leptonic final state are identical to
the $H\,\to\, W^+W^-$ signal.
When trying to determine Higgs boson couplings,
the  $qq\,\to\,qq\,W^+W^-$ cross section must be known precisely, which is
achieved by calculating the next-to-leading order (NLO) QCD corrections.
Such a calculation becomes more crucial when one contemplates using
weak-boson scattering processes, and, more precisely, the absence of strong 
enhancements in these cross sections, as a probe for the existence of a
light Higgs boson~\cite{Bagger:1995mk,Chanowitz:2004gk}. 
Here the knowledge of NLO 
QCD corrections is essential in order to distinguish the enhancement
from strong weak-boson scattering from possible enhancements due to
higher order QCD effects. 

In two recent papers, the calculation of the NLO QCD corrections was presented 
for two simpler vector-boson-fusion processes: the $Hjj$ signal cross 
section~\cite{Figy:2003nv} and the cross sections for $Zjj$
and $Wjj$ production~\cite{Oleari:2003tc}. Both calculations were
turned into fully-flexible parton-level Monte Carlo 
programs.
We here extend this work and describe the calculation and first results for
the NLO QCD corrections to $\wwjj$ production via VBF.  

Weak-boson scattering was first considered in the framework of the
effective $W$ approximation, where the incoming weak bosons are treated
as on-shell particles~\cite{EWA}. This approximation does not provide a
reliable prediction for the kinematical distributions of the forward and
backward jets which are the main characteristic of vector boson fusion
processes~\cite{jet_tag}. 
Calculations of the full $qq\,\to\, qq\,W^+W^-$ processes, first
without $W$ decay~\cite{dic_ww,Baur:1990xe} and then including the full spin 
correlations of the $W$ decay products in the narrow-width
approximation~\cite{Barger:1991ar},  
have been available for more than a decade. Within this
latter approximation, also the real gluon emission contributions, i.e.\ the
${\cal O}(\alpha^4 \alpha_s)$ cross sections for the $pp\,\to\, W^+W^-jjj$
subprocesses, with full spin correlations of the $W$ decay leptons, were
determined~\cite{Duff:1993ut}. 
Very recently, a partonic-level Monte Carlo for all the processes $q_1 q_2
\,\to\, q_3 q_4 q_5 q_6 l \nu$, with exact matrix elements at
${\cal O}(\alpha^6)$, has become available~\cite{Accomando:2005hz}.

\begin{figure}[!htb] 
\centerline{ 
\epsfig{figure=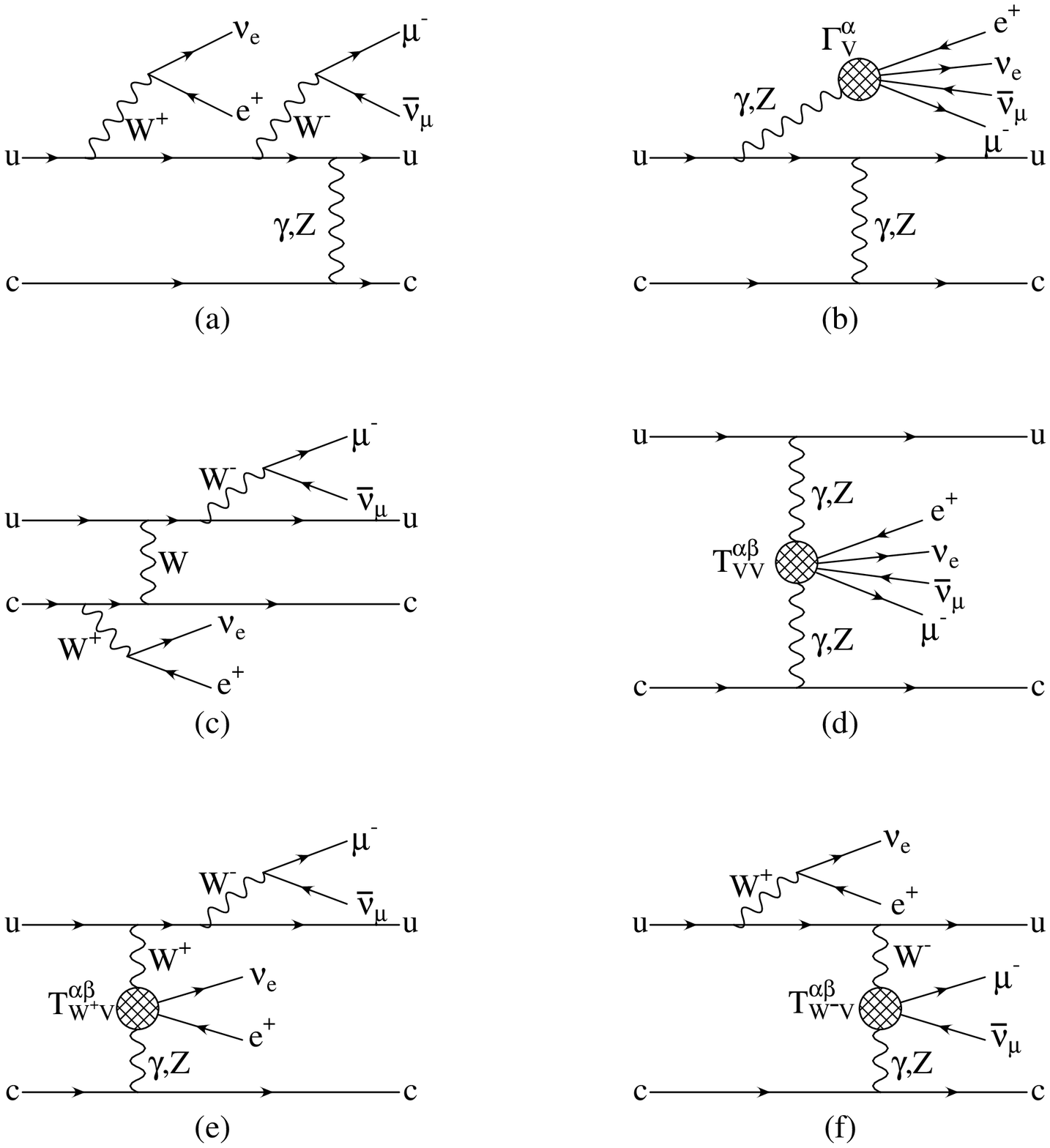,width=0.9\textwidth,clip=} 
} 
\ccaption{} 
{\label{fig:feynBorn} 
The six Feynman-graph topologies contributing to the Born process $uc\,\to\,
uc\,\wwdecay$, a template for neutral-current processes. Diagrams analogous
to (a), (b), (e) and (f), with vector-boson emission off the lower quark
line, are not shown.
}
\end{figure}

In this paper, we consider the proton-proton scattering process 
$pp\,\to\, \wwdecay \, jj(j)\, X$, with all resonant and non-resonant Feynman diagrams and spin
correlations of the final-state leptons included, at order
$\alpha^6\alpha_s$. Since this process is very difficult to detect above
QCD backgrounds, except in phase-space regions which are completely
dominated by $t$-channel electroweak (EW) boson exchange, we only
consider $t$-channel contributions, as explained in
Sec.~\ref{sec:tree_level}.  
In the rest of the paper, we will refer also to
this approximated process as EW $\wwjj$ production.
Electroweak gauge invariance requires that, beyond vector-boson
scattering graphs, also the direct emission of the produced (virtual) 
$W$s off the quark lines be considered. Several examples are depicted in
Fig.~\ref{fig:feynBorn}, which shows the basic Feynman-graph topologies 
which need to be considered for our calculation at tree level, for the 
particular subprocess $uc\,\to\, uc \, \wwdecay$. Real
emission contributions (including quark-gluon initiated subprocesses) are
generated by attaching an external gluon in all possible ways on the two
quark lines in Fig.~\ref{fig:feynBorn}. For the virtual corrections, we
only need to consider Feynman graphs with a virtual gluon attached to a
single quark line: gluon exchange between the up- and the charm-quark
line leads to a color-octet state for the external $u\bar u$ or $c\bar
c$ pair, which cannot interfere with the color-singlet structure at tree
level. As a result, the virtual contributions  contain, at most, pentagon  
diagrams, which arise e.g.\ by connecting the incoming and the outgoing
up-quark in Fig.~\ref{fig:feynBorn}~(a) with a virtual gluon. The other
graphs in Fig.~\ref{fig:feynBorn} lead to box, vertex, or quark
self-energy corrections, and these latter classes have already been 
encountered in Ref.~\cite{Oleari:2003tc}. 

Many aspects of the present calculation parallel this previous work.
The cancellation of collinear and soft divergences for generic VBF processes
was described in detail in Ref.~\cite{Figy:2003nv} and need not be repeated
here, since it can be applied verbatim for the case at hand. The calculation
of vertex and box corrections was needed for the case of $W$ and $Z$
production~\cite{Oleari:2003tc} already, and, thus, these aspects of the
virtual corrections need a brief review only. This review is provided in
Sec.~\ref{sec:calculation}, where we describe the details of our calculation
and the approximations with regard to crossed diagrams in the presence of
identical quark flavors.
As in the previous work, we regularize the loop
integrals via dimensional reduction and separate the virtual
amplitudes into $1/\ep^2$ and  $1/\ep$ terms, which
multiply the Born amplitude, and remaining finite terms, which are then
calculated numerically, using the helicity-amplitude techniques of
Ref.~\cite{HZ}. A major concern here is the numerically stable
and fast evaluation of the pentagon graphs. We make use of Ward
identities and map large fractions of the pentagon contributions  
onto more easily evaluable four-point functions. Another important
feature is the systematic use of ``leptonic tensors'' which describe
groups of purely electroweak subdiagrams. 
%
%

In Section~\ref{sec:MC}, we describe the numerous consistency tests
which we have
performed, ranging from comparison to code generated by
\MADGRAPH~\cite{madgraph} for the tree-level amplitudes to gauge
invariance tests.  In addition, we present the properties of our numerical
Monte Carlo program and how we have dealt with the gauge invariant handling
of finite $W/Z$ widths, the singularities for incoming photons and the choice
of physical parameters.
We then use this Monte Carlo program to produce first results for EW $\wwjj$
production at the LHC. Of particular concern is the scale dependence of the
NLO results, which provides an estimate for the residual theoretical error of
our cross-section calculations. We discuss the scale dependence and the size
of the radiative corrections for various distributions in
Sec.~\ref{sec:pheno}. Conclusions are given in Sec.~\ref{sec:summary}.

\section{Elements of the calculation}
\label{sec:calculation}
Our goal is the calculation of EW $W^+W^-jj$ production cross sections with
NLO QCD accuracy in phase-space regions which are typical for vector-boson
fusion. This implies that some electroweak
contributions, like triple gauge 
boson production ($pp\to W^+W^-V$ with $V\to jj$), can safely be
neglected. These approximations will be specified below.  Also, we make use
of the general structure of NLO QCD corrections to VBF processes: it is
sufficient to specify the contributions to the Born, the real-radiation and
virtual amplitudes which enter the cross section expressions of
Ref.~\cite{Figy:2003nv}.  In this section, we describe how these
contributions have been computed, the approximations used throughout this
calculation and some technical details.

\subsection{Tree-level contribution and approximations}
\label{sec:tree_level}
The Feynman diagrams contributing to
$pp\,\to\, jj\,\wwdecay$, where both resonant and non-resonant
processes are fully considered, can be grouped into six
separate classes. The first group of two, which consists of the VBF
processes considered in this paper, is characterized by $t$-channel 
neutral-current (NC) and charged-current (CC) exchange between the two
scattering quark lines. The other four classes correspond to $u$- and
$s$-channel exchange. The NC and CC labels are assigned depending on the
external quark flavors: the incoming and outgoing
quark charges on each quark line coincide for a neutral current process
and differ by one unit of $|e|$ for a charged current process.

For each neutral-current process, and in the unitary gauge which we use
throughout, there are 181 Feynman graphs, which can 
be grouped into six distinct topologies. Generic diagrams for each of the 
topologies (a) to (f) are shown in 
Fig.~\ref{fig:feynBorn} for the specific 
subprocess $uc\,\to\, uc\,\wwdecay$. They correspond to the following
configurations:
\begin{itemize}
\item[(a)] Two virtual $W$ bosons are emitted 
from the same quark line and in turn decay leptonically. 
%
\item[(b)] A virtual $\gamma$ or $Z$ boson ($V$) with subsequent leptonic
decay is emitted from either quark line.  The tree-level expression for the
sub-amplitude $V\,\to\, \wwdecay$ is given by the tensor $\Gamma_V^\alpha$,
where $\alpha$ is the tensor index carried by the vector boson.
\item[(c)] The leptonically-decaying $W$ bosons are emitted from two
different quark lines.
\item[(d)] Vector-boson fusion in the $t$-channel gives rise to the
sub-amplitude $VV\,\to\, \wwdecay$, which is characterized by the tensor
$T_{VV}^{\alpha\beta}$.  The tensor indices of the
scattering $V$ bosons are indicated with $\alpha$ and $\beta$.
\item[(e)] The leptons are produced by an external $W^-$ boson emitted from a
quark line and a $W^+ V \,\to\,e^+\nu_e$ fusion process in the $t$-channel.
The latter is described by $T_{W^+V}^{\alpha\beta}$.
\item[(f)] The leptons stem from $W^+$ emission from a quark line, 
accompanied by $t$-channel 
$W^- V \,\to\,\mu^-\bar\nu_\mu$ scattering, described by  
$T_{W^-V}^{\alpha\beta}$.
\end{itemize}
%
The propagator factors $1/(q^2-m_V^2+im_V\Gamma_V)$ are included in the
definitions of the sub-amplitudes introduced above, which we call ``leptonic
tensors'' in the following.

\begin{figure}[!htb] 
\centerline{ 
\epsfig{figure=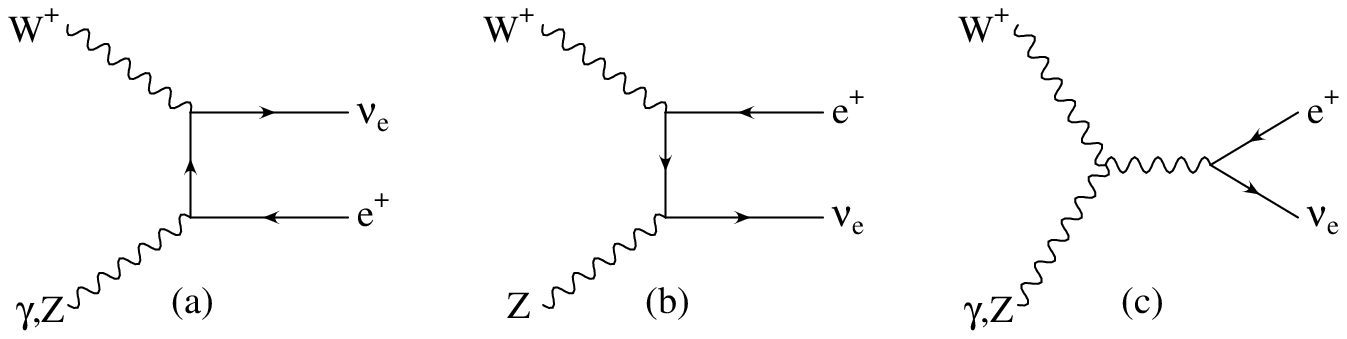,width=0.9\textwidth,clip=} 
} 
\ccaption{} 
{\label{fig:T_WV} 
Diagrams contributing to the scattering amplitude $T_{W^+V}^{\al\beta}$,
depicted in Fig.~\ref{fig:feynBorn}~(e), that describe the tree-level
subprocess $W^+V\to e^+\nu_e $, where $V$ is a $\gamma$ or a $Z$ vector
boson, and $\al$ and $\beta$ are the tensor
indices carried by the charged and neutral vector bosons, respectively.  }
\end{figure} 
The explicit structure of one of these leptonic tensors is
given in Fig.~\ref{fig:T_WV}, where we have plotted the Feynman diagrams
contributing to  $T_{W^+V}^{\al\beta}$:  
a virtual $W^+$ and a virtual $\gamma$ or $Z$ fuse
into a final state $e^+\nu_e$ lepton pair, and the sub-amplitude
corresponding to these three graphs is the leptonic tensor
$T^{\alpha\beta}_{W^+V}$ which appears in graphs like
Fig.~\ref{fig:feynBorn}~(e). 

\begin{figure}[!htb] 
\centerline{ 
\epsfig{figure=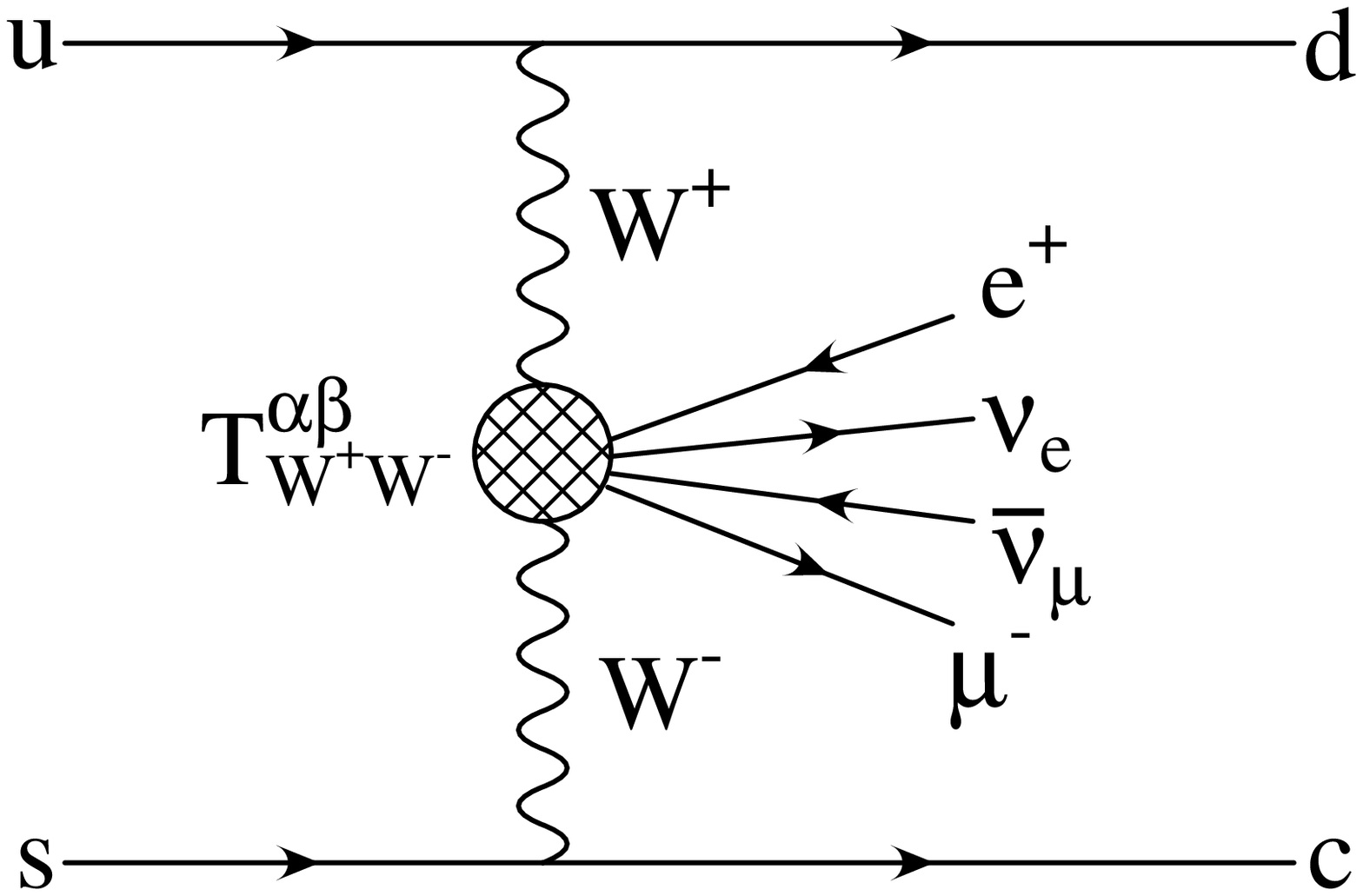,width=0.35\textwidth,clip=} 
} 
\ccaption{} 
{\label{fig:T_WW} 
Contribution from $W^+W^-$ fusion to the scattering process $us\,\to\, dc\,
\wwdecay$. The tensor $T_{W^+W^-}^{\al\beta}$ contains all the tree-level
diagrams contributing to the process $W^+W^-\,\to\,\wwdecay$,
where $\al$ and $\beta$ are the tensor indices carried by the $W^+$ and $W^-$
vector bosons, respectively.  }
\end{figure}

For each charged-current process, such as $us\,\to\,
dc\, \wwdecay$ or $dc\,\to\,us\, \wwdecay$, there are 92 Feynman graphs. 
The different topologies
are completely analogous to the ones for neutral current processes:
simply interchange the $t$-channel bosons $\gamma,Z\leftrightarrow W$ in
Fig.~\ref{fig:feynBorn}. The only new tensor structure that occurs 
is $T_{W^+W^-}^{\al\beta}$, which describes the sub-amplitude for
$W^+W^-\,\to\,\wwdecay$. The corresponding Feynman graph topology is
depicted in Fig.~\ref{fig:T_WW}.

By crossing the external quark lines, one either obtains anti-quark
initiated $t$-channel processes like $\bar uc\to \bar uc\,\wwdecay$ 
(which we fully take into account in our calculation)
or one arrives at NC or CC $s$- or $u$-channel exchange between the two quark
lines, which we count as the other four classes of $jj\wwdecay$ processes:
\begin{itemize}
\item[-] $s$-channel exchange leads to diagrams where all the 
  virtual vector bosons are time-like. They
  correspond to diagrams called conversion, Abelian and
  non-Abelian annihilation in Ref.~\cite{Boudjema:1996qg}, and contain 
  vector-boson production with subsequent decay into pairs of fermions.  
\item[-] $u$-channel exchange occurs for 
  diagrams obtained by interchange of identical initial- or 
  final-state (anti)quarks, such as in the 
  $uu\,\to\, uu\, \wwdecay$ subprocess.
\end{itemize}
In our calculation, we have neglected  
contributions from $s$-channel exchange completely. In addition,
any interference effects of $t$-channel and $u$-channel diagrams 
are neglected. 
This is justified because, in the phase-space 
region where VBF can be observed experimentally, with widely-separated quark 
jets of very large invariant mass, the neglected terms are strongly 
suppressed by large momentum transfer in one or more
weak-boson propagators. Color suppression further reduces any interference
terms.
In Ref.~\cite{Oleari:2003tc} we have checked that, for the analogous 
process $pp\,\to\,W/Z\,jj$, the contribution from the two neglected
classes and from interference effects accounts for
less than 0.3\% of the total cross section, at leading order.
Since we expect QCD corrections to the neglected terms to be modest, the
above approximations are fully justified within the accuracy of our NLO
calculation.

\subsection{Real corrections}
The real-emission corrections to EW $\wwjj$ production with a gluon in the
final state are obtained by attaching one gluon to the quark lines in all
possible ways. 
There are 836 graphs in the case of neutral-current
processes and 444 for the charged-current ones.

The contributions with an initial-state gluon are obtained by crossing the
previous diagrams, promoting the final-state gluon as incoming parton, and an
initial-state (anti-)quark as final-state particle.  
We again remove all diagrams where all electroweak boson propagators are
time-like. Such diagrams, for consistency, must be
removed since we have not considered the corresponding Born
contributions, namely the $s$-channel diagrams corresponding to triple
weak-boson production. These diagrams are strongly
suppressed when VBF cuts (see Sec.~\ref{sec:pheno}) are applied to the
final-state jets.

In the regions of phase space where soft and collinear
configurations can occur, we encounter singularities in the phase-space
integrals of the real-emission squared amplitudes.
The regularization of these singularities in the dimensional-regularization
scheme, with space-time dimension $d=4-2\epsilon$, and the
counter-terms which are needed to get finite expressions  within the 
subtraction method, are discussed extensively in the literature
(see, for example,~\cite{CS}).
Since these divergences only depend on the color structure of the external
partons, the subtraction terms encountered for EW $\wwjj$ production are
identical in form to those found for Higgs boson production in
VBF~\cite{Figy:2003nv} and for EW $Vjj$ production~\cite{Oleari:2003tc}.
The integration over the singular counter-terms yields, after
factorization of the parton distribution function, the contribution 
\beq
\label{eq:I}
<\I(\ep)> = |\MB|^2 \frac{\alpha_s(\mu_R)}{2\pi} C_F
\(\frac{4\pi\mu_R^2}{Q^2}\)^\epsilon \Gamma(1+\epsilon)
\lq\frac{2}{\epsilon^2}+\frac{3}{\epsilon}+9-\frac{4}{3}\pi^2\rq\;.
\eeq
Here, the notation of Ref.~\cite{CS}, but adapted to dimensional
reduction, has been used. $\MB$ denotes the amplitude of the
corresponding Born process and $Q^2$ is 
the momentum transfer between the initial and final state quark in
Fig.~\ref{fig:feynBorn}. 
These singular terms are eventually cancelled by the virtual corrections,
when infrared-safe quantities are computed.

\subsection{Virtual corrections}
\label{sec:virtual}

As for the real-radiation cross sections, the divergences that affect 
the virtual gluon contributions depend on the color structure of the 
external partons.
The main difference with $Hjj$ and $Vjj$ production
is that the finite parts of the virtual corrections are
more complicated for the present case, since the previous two
processes only sport vertex and
box corrections, while now we have to deal with pentagon-type
loop integrals.

The QCD corrections to EW $\wwjj$ production appear as two gauge-invariant
subsets, corresponding to gluon emission and reabsorption on either the
upper or the lower fermion line in
Fig.~\ref{fig:feynBorn}.  Due to the color-singlet nature of the exchanged
electroweak bosons, any interference terms of the Born amplitude with
virtual sub-amplitudes with gluons
attached to both the upper and the lower quark lines vanish identically at
order $\alpha_s$. Hence, it is sufficient to consider radiative corrections
to a single quark line only, which we here take as the upper one. Corrections
to the lower fermion line are an exact copy.
We have regularized the virtual corrections in the dimensional reduction
scheme~\cite{DR_citation}: we have performed the Passarino-Veltman
(PV)~\cite{Passarino:1978jh} reduction of the tensor integrals in $d=4-2\ep$
dimensions, while the algebra of the Dirac gamma matrices, of the external
momenta and of the polarization vectors has been performed in $d=4$
dimensions.

We split the virtual corrections into three classes: 
the virtual corrections along a quark line with only one vector
boson attached (e.g.\ diagram~(d) in Fig.~\ref{fig:feynBorn} or diagrams
(a), (b), (e) and (f) when considering corrections to the lower quark line),
the virtual corrections along a quark line with two vector bosons
attached (e.g.\ diagrams (b), (c), (e), (f)), and the virtual 
corrections along a quark line with three vector bosons attached (e.g.\
diagram~(a)).

{\bf I.} The virtual NLO QCD contribution to any tree-level Feynman
amplitude ${\cal M}_B^{(i)}$ which has a single electroweak boson $V_1$
(of momentum $q_1$) attached to a quark line,
\beq 
q(k_1) \,\to\, q(k_2) + V_1(q_1) \,,
\eeq
is factorizable in terms of
the amplitude for the corresponding Born graph
\bq
\label{eq:vertexvirt} 
{\cal M}_V^{(i)} =
{\cal M}_B^{(i)} \, \frac{\alpha_s(\mu_R)}{4\pi} \, C_F
\(\frac{4\pi\mu_R^2}{Q^2}\)^\epsilon \Gamma(1+\epsilon)
\lq-\frac{2}{\epsilon^2}-\frac{3}{\epsilon}+c_{\rm virt} 
 +\ord{\ep}\rq \,.
\eq 
Here $\mu_R$ is the renormalization scale, and the boson virtuality $Q^2 =
-(k_1-k_2)^2$ is the only relevant scale in the process, since the quarks are
assumed to be massless, $k_1^2 = k_2^2=0$.  In dimensional reduction, the
finite contribution $c_{\rm virt}$ is equal to $\pi^2/3-7$ ($c_{\rm
virt}=\pi^2/3-8$ in conventional dimensional regularization).

\begin{figure}[t] 
\centerline{ 
\epsfig{figure=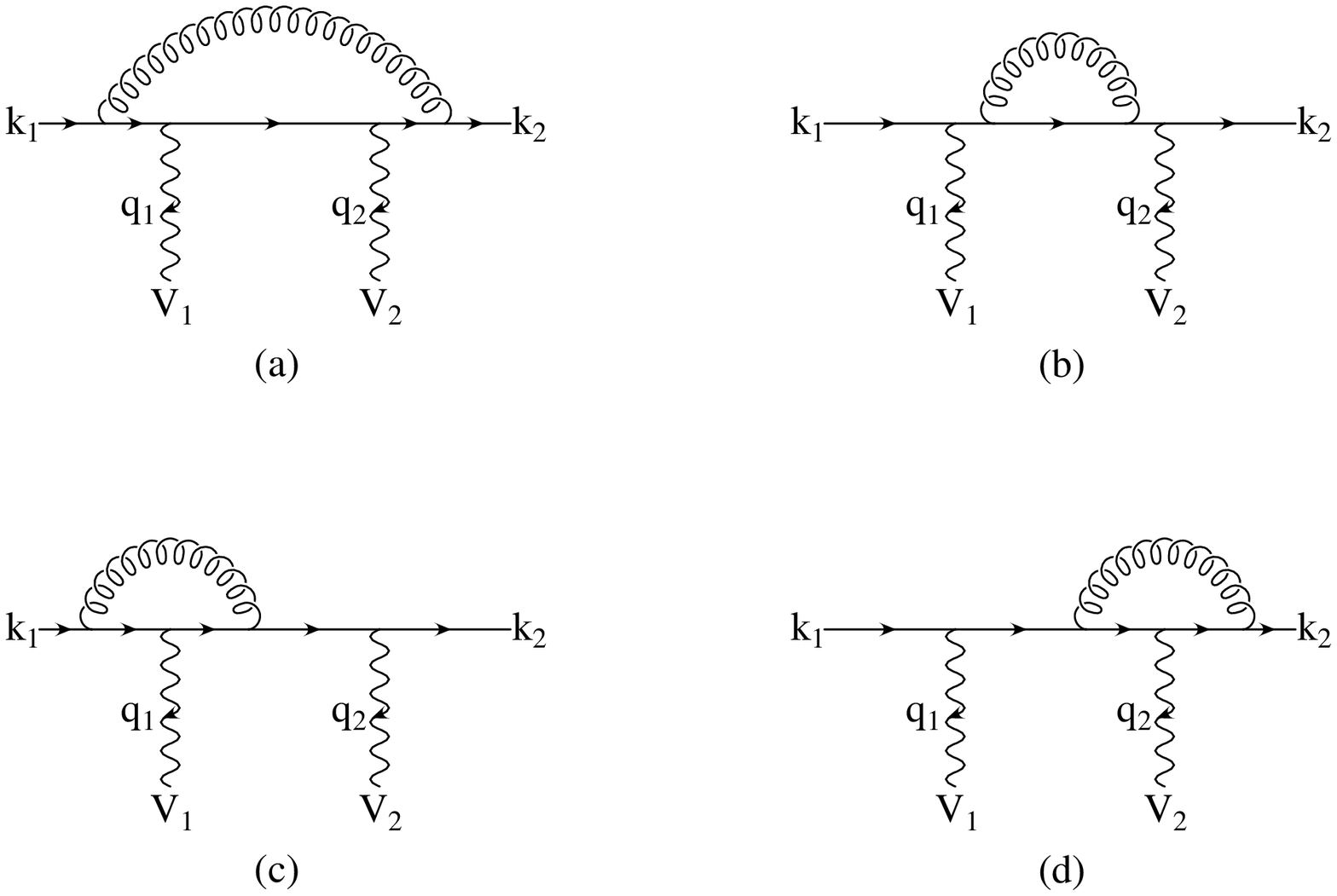,width=0.8\textwidth,clip=} \ \  
} 
\ccaption{} 
{\label{fig:boxline} 
Virtual corrections for a fermion line with two 
attached vector bosons, $V_1(q_1)$ and $V_2(q_2)$. 
The finite part
of the sum of these graphs defines the reduced amplitude 
$\MVV$ of Eq.~(\ref{eq:boxlinefig}).
}
\end{figure} 
{\bf II.} The virtual QCD corrections to the Feynman graphs, where two
electroweak bosons $V_1$ and $V_2$ (of outgoing momenta $q_1$ and $q_2$) are
attached to a quark line, are depicted in Fig.~\ref{fig:boxline}.
It suffices to consider one of the two possible permutations of $V_1$ and
$V_2$, with kinematics
\bq
q(k_1)\,\to \, q(k_2)+V_1(q_1)+V_2(q_2)\,.
\eq
Due to the trivial color structure of the tree-level
diagram, the divergent part (soft and collinear singularities) of the sum of
the four diagrams in Fig.~\ref{fig:boxline} is a multiple of the
corresponding Feynman graph at Born level, 
just like for the vertex corrections,
\beqn
\label{eq:boxlinefig} 
{\cal M}_V^{(i)} &=&
{\cal M}_B^{(i)} \, \frac{\alpha_s(\mu_R)}{4\pi} \, C_F
\(\frac{4\pi\mu_R^2}{Q^2}\)^\epsilon \Gamma(1+\epsilon)
\lq-\frac{2}{\epsilon^2}-\frac{3}{\epsilon}+c_{\rm virt} \rq \nonumber \\
&+&\frac{\alpha_s(\mu_R)}{4\pi} \, C_F \,
\MVVi \, e^2 \,
g_{\tau}^{V_1f_1}g_{\tau}^{V_2f_2}  +{\cal O}(\epsilon) \,.
\eeqn
where we define $Q^2=2\,k_1\cdot k_2$, in order to use the same notation
as in Eq.~(\ref{eq:vertexvirt}).
Here $\tau$ denotes the quark chirality and the electroweak couplings
$g_{\tau}^{Vf}$ follow the notation of Ref.~\cite{HZ}, with, e.g.,
$g_\pm^{\gamma f}=Q_f$, the fermion electric charge in units of $|e|$,
$g_-^{Wf}=1/(\sqrt{2}\sin\theta_W)$ and $g_-^{Zf}=(T_{3f}-Q_f\sin^2\theta_W)/
(\sin\theta_W\cos\theta_W)$,
where $\theta_W$ is the weak mixing angle and $T_{3f}$ is the third component
of the isospin of the (left-handed) fermions.

A finite contribution of the virtual diagrams, which is proportional to the
Born amplitude (the $c_{\rm virt}$ term), is pulled out in correspondence
with Eq.~(\ref{eq:vertexvirt}). The remaining non-universal term, $\MVVi$, is
also finite and can be expressed in terms of the finite parts of the
Passarino-Veltman $B_{ij}$, $C_{ij}$ and $D_{ij}$ functions. The
corresponding analytic expressions were given in
Ref.~\cite{Oleari:2003tc}. Note that the effective polarization vectors
for the electroweak bosons $V_1$ and $V_2$, which enter the expressions
for the $\MVVi$, are $W^\pm$ decay currents, the leptonic tensors
$\Gamma_V^\mu$ (for Fig.~\ref{fig:feynBorn}~(b)) and/or the entire lower
parts of the Feynman graphs for Fig.~\ref{fig:feynBorn}~(c,e,f), when
combining Feynman graphs with identical topology.

\begin{figure}[t] 
\centerline{ 
\epsfig{figure=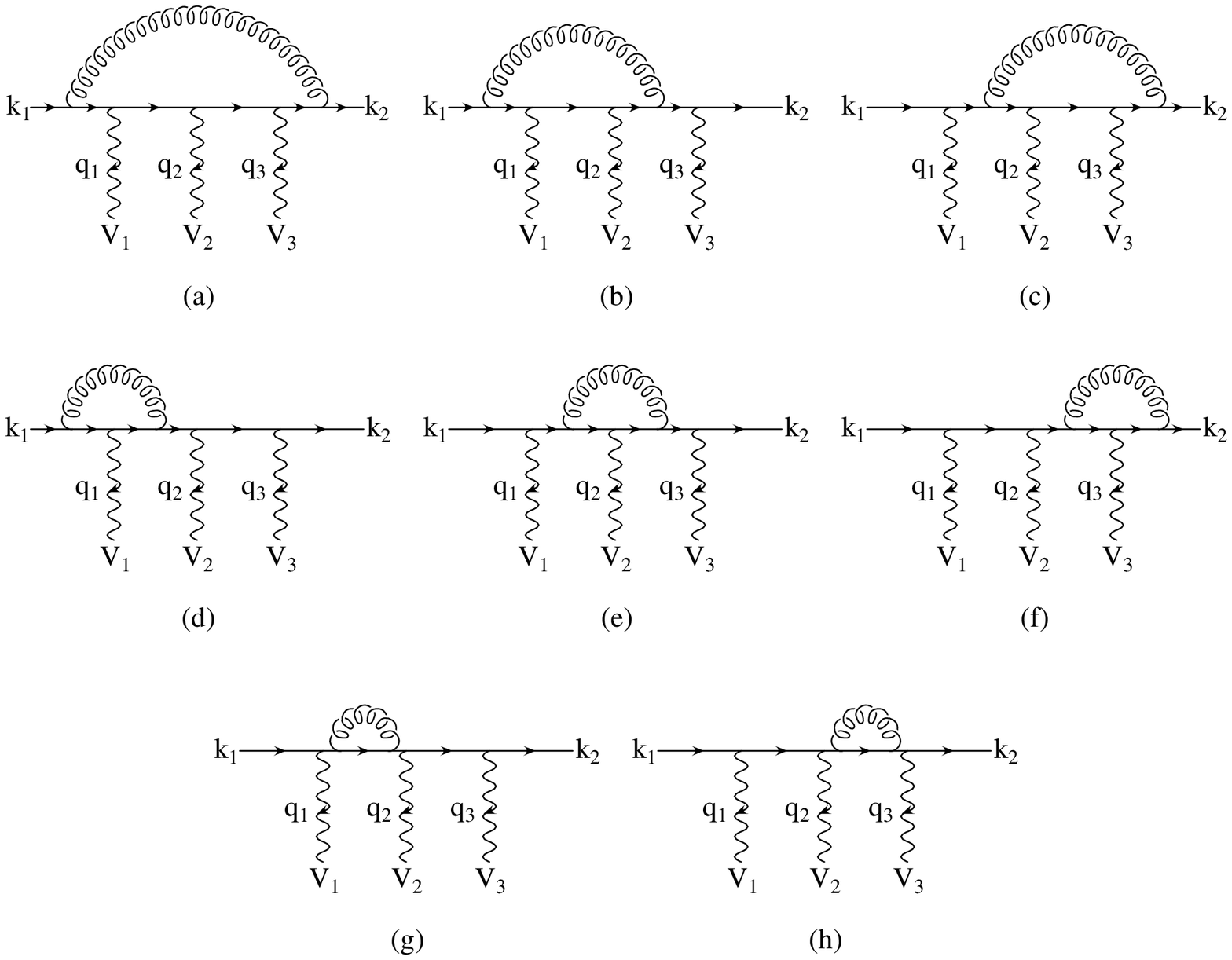,width=0.9\textwidth,clip=} \ \  
} 
\ccaption{} 
{\label{fig:pentaline} Virtual corrections for a fermion line with three
attached vector bosons, $V_1(q_1)$, $V_2(q_2)$ and $V_3(q_3)$. 
The finite part of the sum of these graphs defines the reduced amplitude
$\MVVV$ of Eq.~(\ref{eq:pentalinefig}).  }
\end{figure}
{\bf III.} The virtual QCD corrections to the 
Feynman graphs where three electroweak
bosons $V_1$, $V_2$ and $V_3$ (of outgoing momenta $q_1$, $q_2$ and $q_3$)
are attached to a quark line, are depicted in Fig.~\ref{fig:pentaline}.
It suffices to consider one of the six possible permutations of $V_1$,
$V_2$ and $V_3$, with kinematics 
\beq
q(k_1)\,\to \, q(k_2)+V_1(q_1)+V_2(q_2)+V_3(q_3)\,.
\eeq
The trivial color structure of the tree-level diagram allows the
factorization of the divergent part of the sum of the eight
diagrams in Fig.~\ref{fig:pentaline} in terms of the corresponding Born
sub-amplitude
\beqn
\label{eq:pentalinefig} 
{\cal M}_V^{(i)} &=&
{\cal M}_B^{(i)} \, \frac{\alpha_s(\mu_R)}{4\pi} \, C_F
\(\frac{4\pi\mu_R^2}{Q^2}\)^\epsilon \Gamma(1+\epsilon)
\lq-\frac{2}{\epsilon^2}-\frac{3}{\epsilon}+c_{\rm virt} \rq \nonumber \\
&+&\frac{\alpha_s(\mu_R)}{4\pi} \, C_F \,
\MVVVi \, e^3 \,
g_{\tau}^{V_1f_1}g_{\tau}^{V_2f_2} g_{\tau}^{V_3f_3}  +{\cal O}(\epsilon) \,.
\eeqn
Again, a finite contribution from the virtual diagrams, proportional 
to the Born amplitude ($c_{\rm virt}$), is pulled out and the
remaining finite part is indicated with $\MVVVi$.

The factorization of the divergent parts of the various virtual
contributions, as multiples of the corresponding Feynman amplitudes at
Born level, $\MB^{(i)}$, 
implies that the overall infrared and collinear divergences multiply
the complete Born amplitude, $\MB = \sum_i\MB^{(i)}$.
We can summarize our results for the
virtual corrections to the individual fermion lines by writing the complete 
virtual amplitude ${\cal M}_V$  as 
\beqn
\label{eq:pent_box_tri_contrib}
{\cal M}_V &=& {\cal M}_B\,\frac{\alpha_s(\mu_R)}{4\pi} \, C_F
\(\frac{4\pi\mu_R^2}{Q^2}\)^\epsilon \Gamma(1+\epsilon)
\lq-\frac{2}{\epsilon^2}-\frac{3}{\epsilon}+c_{\rm virt}\rq \nonumber\\
&+&\frac{\alpha_s(\mu_R)}{4\pi} \, C_F \, e^2 \sum_i
\MVVi g_{\tau}^{V_1f_1}g_{\tau}^{V_2f_2} \nonumber\\
&+&\frac{\alpha_s(\mu_R)}{4\pi} \, C_F \, e^3 \sum_i
\MVVVi g_{\tau}^{V_1f_1}g_{\tau}^{V_2f_2} g_{\tau}^{V_3f_3} 
+{\cal O}(\epsilon) \nonumber \\
&=& {\cal M}_B\,\frac{\alpha_s(\mu_R)}{4\pi} \, C_F
\(\frac{4\pi\mu_R^2}{Q^2}\)^\epsilon \Gamma(1+\epsilon)
\lq-\frac{2}{\epsilon^2}-\frac{3}{\epsilon}+c_{\rm virt}\rq 
+\widetilde{\cal M}_V \,,
\eeqn
where the sums run over the different orderings of the attached 
weak bosons and the relevant topologies of Fig.~\ref{fig:feynBorn},
when using effective polarization vectors for the electroweak bosons, as
discussed below Eq.~(\ref{eq:boxlinefig}).
Note that $\widetilde{\cal M}_V$ is completely finite. The NLO
contribution to the cross section at order $\as$ comes from the
interference of the virtual amplitude with the Born term. For
corrections to a quark line it is given by 
\beq    
\label{eq:virtual_born}
2 \Re \lq {\cal M}_V\MB^* \rq
= |\MB|^2 \frac{\alpha_s(\mu_R)}{2\pi} C_F
\(\frac{4\pi\mu_R^2}{Q^2}\)^\epsilon \Gamma(1+\epsilon)
\lq-\frac{2}{\epsilon^2}-\frac{3}{\epsilon}+c_{\rm virt}\rq\
+2 \Re \lq \widetilde{\cal M}_V\MB^* \rq \,.
\eeq
The divergent piece appears as a multiple of the Born amplitude
squared and it cancels explicitly 
against the phase-space integral of the dipole terms (see Ref.~\cite{CS}
and Eq.~(2.10) of Ref.~\cite{Figy:2003nv})
\beq
<\I(\ep)> = |\MB|^2 \frac{\alpha_s(\mu_R)}{2\pi} C_F
\(\frac{4\pi\mu_R^2}{Q^2}\)^\epsilon \Gamma(1+\epsilon)
\lq\frac{2}{\epsilon^2}+\frac{3}{\epsilon}+9-\frac{4}{3}\pi^2\rq\,,
\eeq
which absorbs the real-emission singularities which are left after
factorization of the parton distribution functions. After this
cancellation, all the remaining integrals are finite and can, hence, be
evaluated in $d=4$ dimensions.

\subsection{Technical details}

Our Monte Carlo program computes all amplitudes numerically, using the
helicity technique and the formalism of Ref.~\cite{HZ}.
For the tree-level and real-emission amplitudes (including
counter-terms), the method is straightforward, since these contributions are
finite at each phase-space point.
The evaluation of the helicity amplitudes is very fast, due to 
the modular structure that one achieves by grouping the whole set of
diagrams according to the topologies illustrated in
Fig.~\ref{fig:feynBorn}.
The $W^+\to e^+\nu_e$ and $W^-\to\mu^-\bar\nu_\mu$ decay amplitudes and
the single index leptonic tensors $\Gamma_V^\alpha$ ($V=Z,\;\gamma$) 
are effective polarization vectors which only depend on the lepton
momenta. They are the same for all subprocesses, i.e.\ they do not depend
on quark flavor or whether quarks and/or anti-quarks scatter. Similarly
the second-rank leptonic tensors $T_{VV}^{\alpha\beta}$,
$T_{W^\pm V}^{\alpha\beta}$ and $T_{W^+W^-}^{\alpha\beta}$ are
independent of quark flavor and come in just two kinematic
configurations, depending on whether or not 
an external gluon is attached to the upper or the lower quark
line in Fig.~\ref{fig:feynBorn}. Correspondingly, the leptonic tensors
are calculated first in our numerical program and then used in crossed
subprocesses and subtraction terms. 
The code for these leptonic tensors has been generated
with \MADGRAPH\ and adapted to the tensor structure
required for our full program. We note, in passing, that this approach
allows for 
straightforward inclusion of new physics effects in the electroweak
sector: only the leptonic tensors would be affected by modifications
like anomalous three- or four-gauge-boson couplings or strong
electroweak-boson scattering. 
One major advantage of the modular strategy is the increase in
computational speed. In the calculation of the
real-emission contributions, which constitute the most CPU-time
intensive part of the code, our program is about 70 times faster than a
direct use of \MADGRAPH-generated routines for the individual
subprocesses.

Special care has to be taken in the extraction of the finite parts $\MVV$ and
$\MVVV$, which are contained in the full virtual amplitude of
Eq.~(\ref{eq:pent_box_tri_contrib}).  In order to keep the expressions small
and fast to evaluate, we have implemented the PV tensor reduction
numerically.  Here we are adopting a natural extension of the PV notation,
and we call $E_{ij}$ the coefficient functions from the tensor reduction of
pentagon integrals.  
Since the finite $\MVV$ and $\MVVV$ virtual sub-amplitudes
only contain the finite pieces of
the various tensor integrals, one needs to track how the divergent
contributions in the 
expressions of the scalar integrals feed into the expressions of the tensor
coefficients $B_{ij}$, $C_{ij}$, $D_{ij}$ and $E_{ij}$, and how they generate
finite contributions in coefficients that contain a factor $(d-4)$
in the numerator.
The resulting analytical expression for $\MVV$, in terms of finite 
functions, is given in Ref.~\cite{Oleari:2003tc}.
We postpone to a future paper~\cite{future_JOZ} any further technical
discussion about the computation of $\MVVV$.

\section{Checks and implementation in a parton-level Monte Carlo}
\label{sec:MC}

The cross-section contributions discussed in the previous section have been
implemented in a fully-flexible parton-level Monte Carlo, which is very
similar to the programs for $Hjj$ and $Vjj$ production in VBF as described in
Refs.~\cite{Figy:2003nv} and~\cite{Oleari:2003tc}. 
The matrix-element calculation is 
divided into three main parts, that deal with the evaluation of the
tree-level, the real-emission and the virtual contributions. All
elements have been extensively tested as detailed below.\\
\\
{\bf Tree-level contribution}\\
We have compared our tree-level code with purely \MADGRAPH\ generated
output, and we have found agreement with a typical relative accuracy of
$10^{-10}$.\\
\\
{\bf Real-radiation contribution}\\
The same comparison has been performed for the real radiation
contributions, with typical agreement at the $10^{-10}$ level.  In
addition, we have also checked the QCD gauge invariance of the real-emission
corrections.  More specifically, the real-emission amplitude for the process
$ q q' \to q q' g \, \, \wwdecay$ has the form
\beq
\label{eq:qcdgauge}
\mathcal{M}_R=\varepsilon_\mu(p)\,\mathcal{M}^\mu_R \,,
\eeq
where $p$ is the momentum of the emitted gluon and $\varepsilon_\mu(p)$ its
polarization vector.
Gauge invariance demands that the amplitude $\mathcal{M}_R$ remains unchanged
upon the substitution  $\varepsilon_\mu(p) \,\to$
$\varepsilon_\mu(p)+\beta p_{\mu}$ (with $\beta$ arbitrary), that is
\beq
  p_{\mu} \mathcal{M}^\mu_R  = 0.
\eeq
This relation is satisfied within the numerical accuracy of the
program.\\
\\
{\bf Virtual contribution: code checks}\\
As far as the virtual contribution is concerned, we have implemented two
different codes, one analytical, in \MAPLE, and one 
numerical, in \fortran.
The analytical code sums all the eight Feynman diagrams in
Fig.~\ref{fig:pentaline}, which we call $\mathcal{P}_{\mu_1\mu_2\mu_3}$
for uncontracted polarization vectors of the three electroweak bosons, 
and writes it in terms of the PV coefficient functions,
$B_{ij}, \ldots E_{ij}$, in $d$ dimensions. We schematically represent this
tensor reduction by
\beq
\label{eq:PV_P}
 \mathcal{P}_{\mu_1\mu_2\mu_3}(k_1,q_1,q_2,q_3) = 
 \sum_{ij} \, T^{(ij)}_{\mu_1\mu_2\mu_3} \, {\rm (PV)}_{ij} \,,
\eeq
where ${\rm (PV)}_{ij}=\{B_{ij},C_{ij},D_{ij},E_{ij}\}$ is one of the
Passarino-Veltman coefficient functions, and the (finite) tensors
$T^{(ij)}_{\mu_1\mu_2\mu_3}$ correspond to spinor products describing
the quark lines in Fig.~\ref{fig:pentaline}.  The
$\mathcal{P}_{\mu_1\mu_2\mu_3}$ and the $(PV)_{ij}$ still contain
divergent contributions. We denote 
their finite parts by $\tilde{\mathcal{P}}_{\mu_1\mu_2\mu_3}$,
$\tilde{B}_{ij},\;\tilde{C}_{ij},\;\tilde{D}_{ij},\;\tilde{E}_{ij}$,
respectively.

The analytic code contains all the recursion relations that can be
used to 
reduce the PV coefficient functions to combinations of scalar integrals only:
$B_0, C_0, D_0$ and $E_0$ functions.  The $E_0$ function can be further
expressed in terms of the sum of five $D_0$ functions, as described in
Ref.~\cite{BDK}, when $d=4-2\ep$, in the limit $\ep \,\to\, 0$. The
analytic continuation of $D_0$ functions was checked against
Ref.~\cite{Dupl:2001}. 
The tensor reduction down to scalar integrals, and the direct substitution of
the corresponding expressions computed in $d=4-2\ep$ dimensions have
been used to check the structure of the divergent terms, and to show that,
once contracted with the Born amplitude, they are given by
Eq.~(\ref{eq:pent_box_tri_contrib}).

The expression of $\tilde{\mathcal{P}}_{\mu_1\mu_2\mu_3}$ in terms of 
PV coefficient
functions is turned by \MAPLE\ into a \fortran\ code, where care is taken to
obtain the correct limit when $d \,\to\, 4$.  All the technical details about
this part of the program will be given in Ref.~\cite{future_JOZ}.

\begin{figure}[t] 
\centerline{ 
\epsfig{figure=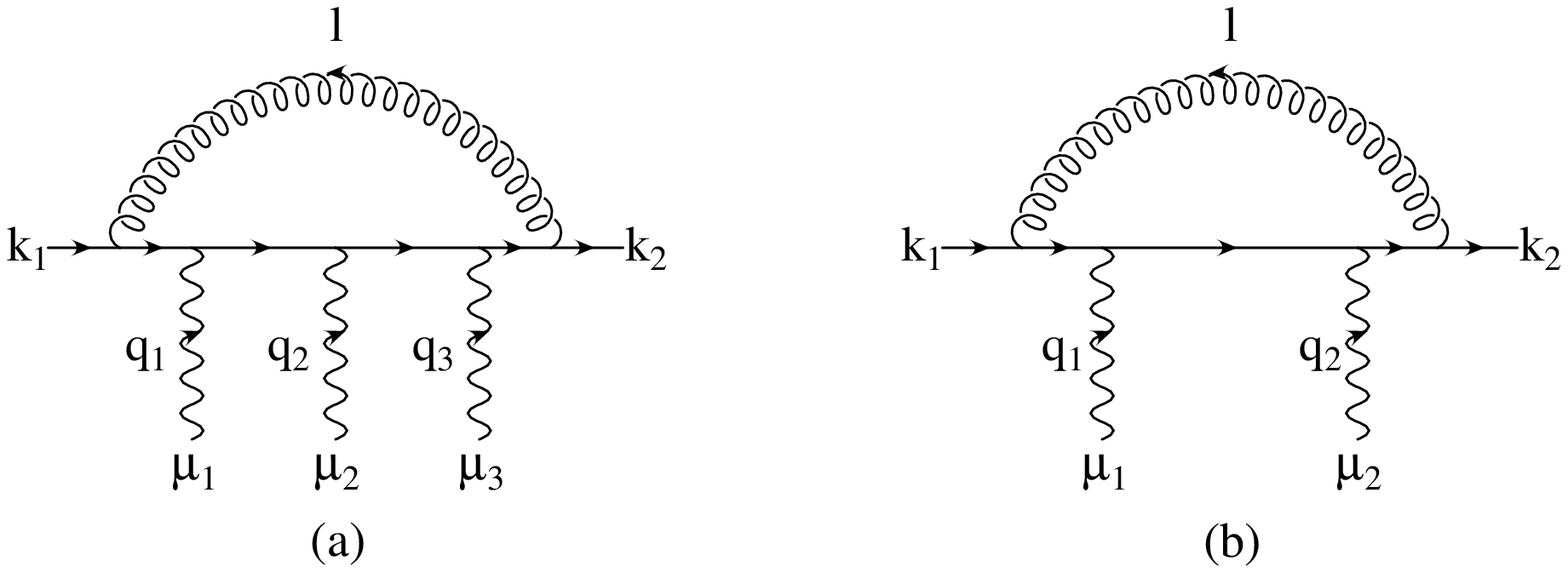,width=0.8\textwidth,clip=} \ \  
} 
\ccaption{} 
{\label{fig:E_munuro} $\mathcal{E}_{\mu_1\mu_2\mu_3}(k_1,q_1,q_2,q_3)$ and
$\mathcal{D}_{\mu_1\mu_2}(k_1,q_1,q_2)$ of
Eqs.~(\protect{\ref{eq:pline}})--(\protect{\ref{eq:pent_contract_3}}).} 
\end{figure}
Both the analytical and the \fortran\ code have been checked extensively
using gauge invariance, applied at different levels of complication: 
\begin{itemize}
\item[-]
at the
level of the single pentagon loop (diagram (a) of Fig.~\ref{fig:E_munuro}),
\item[-]
at the level of the sum of all the virtual corrections along a single quark
line, $\mathcal{P}_{\mu_1\mu_2\mu_3}$, 
\item[-] and at the level of the entire
scattering process (for the \fortran\ code).  
\end{itemize}
To illustrate an example of gauge check, we consider the simpler case of the
pentagon loop of diagram~(a) in Fig.~\ref{fig:E_munuro}
\beq
\label{eq:pline}
\mathcal{E}_{\mu_1\mu_2\mu_3}(k_1,q_1,q_2,q_3) \equiv
\!\!\int\!\!\frac{d^dl}{(2\pi)^d}  
\gamma^\al
\frac{1}{\sla l + \sla k_1 + \sla q_{123}} \gamma_{\mu_3}
\frac{1}{\sla l + \sla k_1 + \sla q_{12}} \gamma_{\mu_2}
\frac{1}{\sla l + \sla k_1 + \sla q_{1}} \gamma_{\mu_1}
\frac{1}{\sla l + \sla k_1 } \gamma_\al
\frac{1}{l^2}\,,
\eeq
where $q_{12}= q_1 + q_2$, $q_{123}=q_1+q_2+q_3$, $k_2=q_{123}+k_1$.
Gauge invariance is simply the statement that, 
upon contracting any one tensor index with the corresponding momentum, and
expressing the contracted gamma matrix as the difference of the two adjacent
fermionic propagators, the pentagon can be reduced to box integrals
$\mathcal{D}_{\mu\nu}$ (see diagram (b) in Fig.~\ref{fig:E_munuro})
\beqn
\label{eq:pent_contract_1}
q_1^{\mu_1} \mathcal{E}_{\mu_1\mu_2\mu_3}(k_1,q_1,q_2,q_3) &=&
\mathcal{D}_{\mu_2\mu_3}(k_1,q_1+q_2,q_3) -
\mathcal{D}_{\mu_2\mu_3}(k_1+q_1,q_2,q_3)\,,\\ 
q_2^{\mu_2} \mathcal{E}_{\mu_1\mu_2\mu_3}(k_1,q_1,q_2,q_3) &=&
\mathcal{D}_{\mu_1\mu_3}(k_1,q_1,q_2+q_3) -
\mathcal{D}_{\mu_1\mu_3}(k_1,q_1+q_2,q_3)\,,\\
\label{eq:pent_contract_3}
q_3^{\mu_3} \mathcal{E}_{\mu_1\mu_2\mu_3}(k_1,q_1,q_2,q_3) &=&
\mathcal{D}_{\mu_1\mu_2}(k_1,q_1,q_2) -
\mathcal{D}_{\mu_1\mu_2}(k_1,q_1,q_2+q_3)\,. 
\eeqn
%
%
%
Using the PV tensor reduction, we can express $\mathcal{E}_{\mu_1\mu_2\mu_3}$
as a sum of coefficient functions up to $E_{ij}$ (see
Eq.~(\ref{eq:PV_P})), and  
$\mathcal{D}_{\mu\nu}$ as sum of coefficient functions up to $D_{ij}$, and
generate the corresponding \fortran\ code for their finite parts.
We can then check that the analytic or numeric  expression for 
$\tilde{\mathcal{E}}_{\mu_1\mu_2\mu_3}$, once an external index is contracted 
with the corresponding momentum,  agrees with the right-hand-sides of
Eqs.~(\ref{eq:pent_contract_1})--(\ref{eq:pent_contract_3}). Analogous
relations hold at the level of the
$\tilde{\mathcal{P}}_{\mu_1\mu_2\mu_3}$ which
represent the sum of all the virtual corrections along a single quark
line. Both tests are a strong check on the correctness of the entire code.

Finally, we have implemented two independent codes to compute the virtual
corrections for the neutral-current contributions. The relative amplitudes
agree within the numerical precision of the two \fortran\ programs.\\
\\
{\bf Virtual contribution: numerical stability}\\
Gauge invariance has been used not only to check the entire code but is used
every time that a virtual contribution is computed at a given point in 
phase space.
When the diagrams of 
Fig.~\ref{fig:pentaline} are contracted with the leptonic currents which
represent the $W^+\to e^+\nu_e$ and $W^-\to\mu^-\bar\nu_\mu$ decay
amplitudes, the helicity amplitude has the generic form
\beq
J_1^{\mu_1}J_2^{\mu_2} 
\tilde{\mathcal{P}}_{\mu_1\mu_2\mu_3}\,.
\eeq
For example, in the computation of the virtual corrections for the
sub-amplitude~(a) in Fig.~\ref{fig:feynBorn}, we need to evaluate
\beq
\label{eq:JJJpent}
J_+^{\mu_1}J_-^{\mu_2} 
\tilde{\mathcal{P}}_{\mu_1\mu_2\mu_3}(k_1,q_{+},q_{-},q_{0})\,,  
\eeq
where $J_+$ is the electronic current from the decay of a $W^+$ 
with incoming momentum $q_+$, $J_-$ is the
muonic current from the decay of a $W^-$ 
with incoming momentum $q_-$ and $q_{0}$ is the incoming
momentum of the neutral vector boson. 
We  evaluate this expression by projecting the four-vectors
$J_{\pm}$ on the respective momenta
\beq
\label{eq:projectJ}
  J_{\pm}^\mu = x_{\pm}\, q_{\pm}^\mu + r_{\pm}^\mu\,,
\eeq
in such a way that, in the center-of-mass system of the $W^+W^-$ pair, 
the vectors $r_{\pm}$ have zero time component
\beq
  r_{\pm} \cdot \(q_+ + q_- \) =0\,,
\eeq
so that
\beq
x_{\pm} = \frac{J_{\pm}\cdot\(q_+ + q_- \) }{q_{\pm}\cdot\(q_+ + q_- \)}\,.
\eeq
Equation~(\ref{eq:JJJpent}) then becomes
\beq
\label{eq:pent_to_box}
J_+^{\mu_1}J_-^{\mu_2}
\tilde{\mathcal{P}}_{\mu_1\mu_2\mu_3}(k_1,q_{+},q_{-},q_{0})  = 
r_+^{\mu_1}\,r_-^{\mu_2}\,
\tilde{\mathcal{P}}_{\mu_1\mu_2\mu_3}(k_1,q_{+},q_{-},q_{0}) + 
{\rm box\  contributions}\,, 
\eeq
where we have used
Eqs.~(\ref{eq:pent_contract_1})--(\ref{eq:pent_contract_3}).

\begin{figure}[thb] 
\centerline{ 
\epsfig{figure=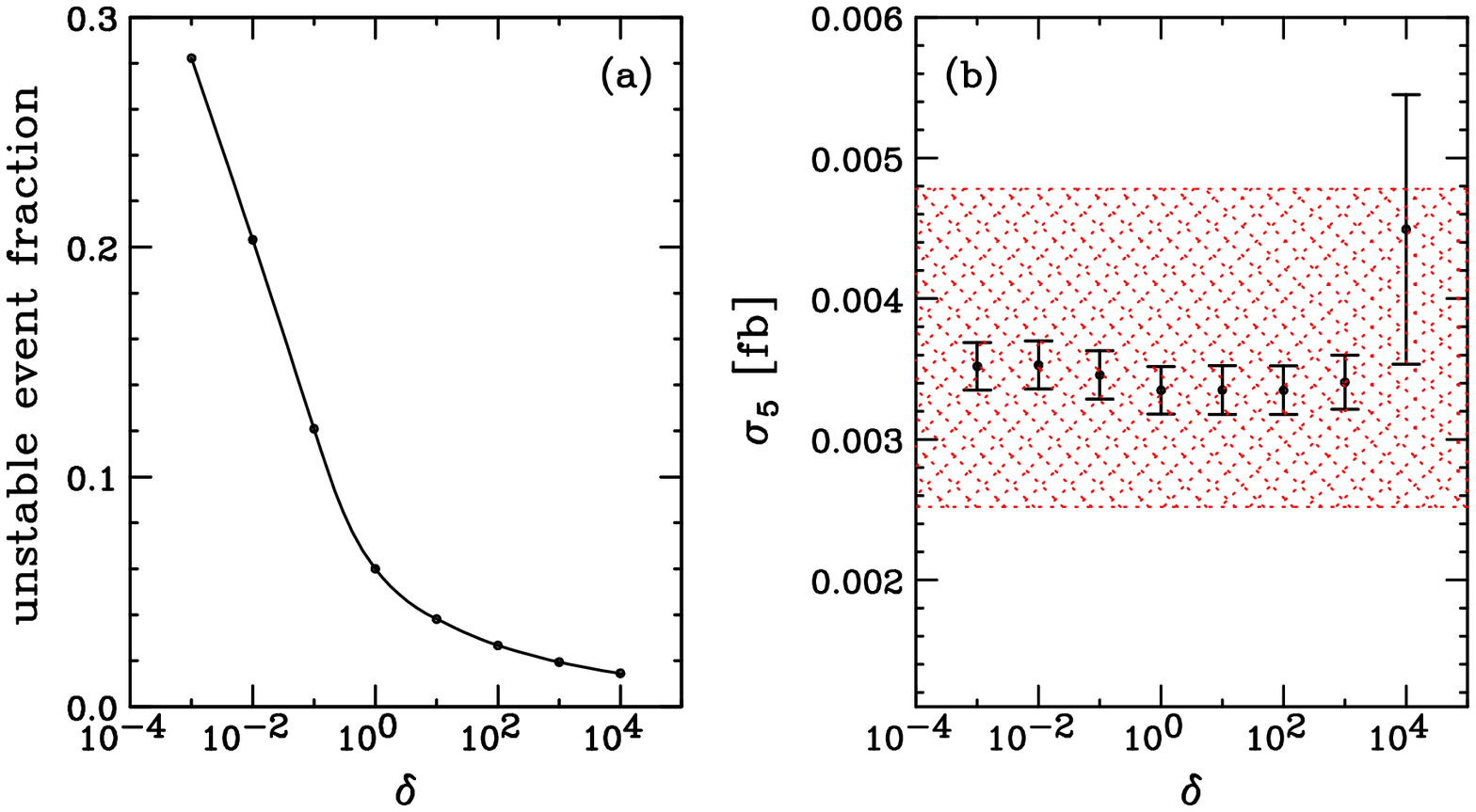,width=0.9\textwidth,clip=} \ \  
} 
\ccaption{} 
{\label{fig:pentagon_error} 
Dependence of the pentagon contribution to the cross section on the
maximal 
relative numerical error, $\delta$, which is allowed for the Ward 
identities of the residual pentagon diagrams. The fraction 
$f$ of subprocess events, where a pentagon subamplitude is discarded for
numerical reasons, is shown in panel~(a).
Panel~(b) gives the pentagon contribution, $\sigma_5$,
to the cross section.
Error bars correspond to the statistical error of the Monte Carlo
integration. 
For comparison, the statistical error on the overall cross section for a
high statistics run with $\pm\, 0.06\%$ accuracy is indicated by the
horizontal band. 
} 
\end{figure}

The projections of Eqs.~(\ref{eq:projectJ})--(\ref{eq:pent_to_box}) 
reduce the magnitude of the coefficients multiplying the pentagon
loops and their overall contribution to the virtual corrections. This
``true'' pentagon contribution to the cross section, defined by the
interference of the residual 
$r_+^{\mu_1}\,r_-^{\mu_2}\, \tilde{\mathcal{P}}_{\mu_1\mu_2\mu_3}$
type terms with the Born amplitude, is called $\sigma_5$ below.
Minimizing it is important in view of the fact that, in the 
tensor-reduction procedure \`a la Passarino-Veltman, Gram determinants 
appear in the denominators of the PV coefficient functions.  There are
points in phase space where these determinants become small and
numerical results become unstable.
We have developed a strategy to interpolate over these critical points (see
Ref.~\cite{future_JOZ}) and, to make sure that the numerical accuracy is not
spoilt, we check,  numerically, that the analogs of
Eqs.~(\ref{eq:pent_contract_1})--(\ref{eq:pent_contract_3}) are
satisfied for the tensors $\tilde{\mathcal{P}}_{\mu_1\mu_2\mu_3}$. In
Fig.~\ref{fig:pentagon_error}~(a) we show the fraction of subprocess
events, $f$ (counting pentagon corrections to the upper and the lower quark
line as different subprocesses), where these Ward identities are violated
by more than a fraction $\delta$. Numerical instabilities of
$\delta=10\%$ or more, for example, affect about $f=13\%$ of the 
generated subprocess events. For subprocess events with Ward identity
violations exceeding $\delta$ we
discard the numerically unreliable $\widetilde{\cal M}_{V_1 V_2 V_3}$ 
and correct the remaining pentagon contributions by a global factor,
$1/(1-f)$, in order to compensate for the loss. As demonstrated in
Fig.~\ref{fig:pentagon_error}~(b), this procedure
leads to a constant overall pentagon contribution, $\sigma_5$, 
when varying $\delta$ between 0.001 and 1000. Numerical
instabilities become large for $\delta \gsim 10^4$. 
For our Monte Carlo runs we choose $\delta=0.1$.
Since the pentagon contribution amounts to less than 0.5\% of the
cross section for EW $\wwjj$ production with VBF cuts, 
the error that is induced by this 
approximation affects our final NLO results at an insignificant
level. For comparison the shaded horizontal band  
shows the size of the overall cross-section uncertainty for a high-statistics
run, with an overall statistical error of 0.06\%. 

The tensor reduction of the box-type virtual contributions is quite
stable, numerically. We have checked that the corresponding
Ward identities, derived in a similar way as for the pentagons, are
violated at only one out of $10^6$ phase-space points by more than
1$\permil$.

The box- and pentagon-type virtual corrections, the finite 
$\widetilde{\cal M}_{V_1 V_2}$ and
$\widetilde{\cal M}_{V_1 V_2 V_3}$ terms in 
Eqs.~(\ref{eq:pent_box_tri_contrib}) and~(\ref{eq:virtual_born}), whose
evaluation is cumbersome and time consuming, amount to less than one
percent of the full
cross section. Therefore, the statistical error of these
contributions affects the accuracy of the full result only marginally and the
number of Monte Carlo events for the computation of the box and pentagon
corrections can be reduced substantially with respect to the Born cross
section and the leading $c_{\rm virt}|\MB|^2$ virtual contribution in
Eq.~(\ref{eq:virtual_born}): in our program, the Monte Carlo statistics
is reduced by a factor 16 for the generic box contributions and by a
factor 128 for the pentagon contributions after the projections of
Eqs.~(\ref{eq:projectJ})--(\ref{eq:pent_to_box}). 
These elements, together with the efficient handling of leptonic
tensors and the other speed-up ``tricks'' described in the previous
section, yield a fast code, which allows us to
perform high-statistics runs with small relative errors on the full NLO cross
sections and distributions. For example, it took about five days of CPU
time on a 3~GHz Pentium~4 PC to obtain an accuracy of 1$\permil$ on the
distributions shown in the next section.

As discussed in detail in Ref.~\cite{Oleari:2003tc}, 
care has to be taken in the
treatment of finite-width effects in massive vector-boson propagators. In
order to handle diagrams where vector bosons decay, like
$W(p_{\ell}+p_{\nu}) \, \to \, \ell(p_{\ell}) +\nul(p_{\nu})$, a finite
vector-boson width $\Gamma_V$ has to be introduced in the resonant poles of
each $s$-channel vector-boson propagator. 
However, in the presence of single- and non-resonant graphs, like (a) and (b)
in Fig.~\ref{fig:T_WV}, this introduces violations of electroweak gauge
invariance in a sub-class of diagrams, which would hold 
in the zero-width approximation.
In the past, different methods, such as the overall factor
scheme~\cite{Baur:1991pp} and the complex-mass scheme~\cite{Denner:1999gp},
have been applied to overcome these problems.  We resort to a modified
version of the complex-mass scheme, which already has been used in
Ref.~\cite{Oleari:2003tc}. We globally replace $m_V^2$ with $m_V^2 -i
m_V\Gamma_V$, while keeping a real value for $\sin^2\theta_W$. This
prescription has the advantage of preserving the electromagnetic Ward
identity which relates the tree-level triple gauge-boson vertex and the
inverse $W$ propagator~\cite{lopez}. It thereby avoids large contributions
from gauge-invariance-violating terms.

Throughout the calculation, fermion masses are set to zero,
because observation of either leptons or (light) quarks in a hadron-collider
environment requires large transverse momenta and hence sizable scattering
angles and relativistic energies.  For consistency, external $b$- and
$t$-quark contributions are excluded.

We have used a diagonal form (equal to the identity matrix) for the
Cabibbo-Kobayashi-Maskawa matrix, $V_{\rm CKM}$. This approximation is not a
limitation of our calculation. As long as no final-state quark flavor is
tagged (no $c$ tagging is done, for example), the sum over all flavors, using
the exact $V_{\rm CKM}$, is equivalent to our results, due to the unitarity of
the $V_{\rm CKM}$ matrix.

The VBF cuts, discussed in Sec.~\ref{sec:pheno}, force the LO differential
cross section for $\wwjj$ to be finite, since they require two
well-separated jets of finite transverse momentum.
For the NLO contributions, initial-state singularities, due to collinear 
$q\,\to\, qg$ and $g\,\to\,
q\bar{q}$ splitting, are factorized into the respective quark and gluon
distribution functions of the proton.
An additional divergence is encountered in those real-emission
diagrams, where a $t$-channel photon of low virtuality is exchanged, thereby
giving rise to a collinear $q \to q\gamma$ singularity. 
We avoid it by imposing a cut on the virtuality of the photon,
$Q_{\gamma,\mathrm{min}}^2=4$~GeV$^2$. Events that do not pass this cut are
considered to be part of the QCD corrections to the
$p\gamma\,\to\,\wwjj$ cross section, that we do not calculate here.

For the computation of cross sections and distributions presented in the
following section, we have adopted the CTEQ6M parton distributions with 
$\alpha_s(m_Z)=0.118$ at NLO, and the CTEQ6L1 set for the LO
calculation~\cite{cteq6}. 
The CTEQ6 parton distributions include $b$ quarks as active flavors. 
However, since in our calculation all fermion masses are neglected,
we have disregarded external $b$- and $t$-quark contributions throughout. 

We have chosen $M_Z=91.188$~GeV, $M_W=80.423$~GeV and $G_F=1.166\times
10^{-5}/$GeV$^2$ as electroweak input parameters. The other parameters,
$\alpha_{\mathrm{QED}}=1/132.54$ and $\sin^2\theta_W=0.22217$, are computed
thereof via LO electroweak relations.  
Final-state partons are recombined into jets according to the $k_T$
algorithm~\cite{kToriginal}, as described in Ref.~\cite{kTrunII}, with
resolution parameter $D=0.8$.

\section{Results for the LHC}
\label{sec:pheno}

The parton-level Monte Carlo program described in the previous section
has been used to determine the size of the NLO QCD corrections to the
EW $\wwjj$  cross sections at the LHC. Using the $k_T$ algorithm, we 
calculate the partonic cross sections for events with at least two 
hard jets,  which are required to have
\beq
\label{eq:cuts1}
p_{Tj} \geq 20~{\rm GeV} \, , \qquad\qquad |y_j| \leq 4.5 \, .
\eeq
Here $y_j$ denotes the rapidity of the (massive) jet momentum which is 
reconstructed as the four-vector sum of massless partons of 
pseudorapidity $|\eta|<5$. The two reconstructed jets of highest transverse 
momentum are called ``tagging jets''. At LO, they are identified with the 
final-state quarks which are characteristic for vector-boson fusion processes. 

We consider the specific leptonic final state $\wwdecay$.
One obtains the cross sections for the phenomenologically more
interesting final state containing any combination of electrons or muons 
($e^+e^-\nu\bar\nu$, $\mu^+\mu^-\nu\bar\nu$, $e^\pm\mu^\mp\nu\bar\nu$, 
but neglecting
identical lepton interference and $ZZ$ final states) by multiplying our
cross sections by a factor of 4.  In order to ensure that the charged 
leptons are well observable, we impose the lepton cuts
\beq
\label{eq:cuts2}
p_{T\ell} \geq 20~{\rm GeV} \,,\qquad |\eta_{\ell}| \leq 2.5  \,,\qquad 
\triangle R_{j\ell} \geq 0.4 \, ,
\eeq
where $\triangle R_{j\ell}$ denotes the jet-lepton separation in the
rapidity-azimuthal angle plane. In addition, the charged leptons are
required to fall between the rapidities of the two tagging jets
\beq
\label{eq:cuts3}
y_{j,min}  < \eta_\ell < y_{j,max} \, .
\eeq
Backgrounds to VBF are significantly suppressed by requiring
a large rapidity separation of the two tagging jets. We here impose the
cut
\beq
\label{eq:cuts4}
\Delta y_{jj}=|y_{j_1}-y_{j_2}|>4\; .
\eeq
Furthermore, we require the two tagging jets to reside in opposite 
detector hemispheres,
\beq
\label{eq:cuts5}
y_{j_1} \times y_{j_2} < 0\, ,
\eeq
with an invariant mass 
\beq
\label{eq:cuts6}
M_{jj} > 600~{\rm GeV}\;.
\eeq
The resulting total cross section receives two major contributions,
arising from the Higgs resonance, via $H\to WW$ decays, and from the
$WW$ continuum, which effectively starts at the $W$-pair threshold.
Already for Higgs boson masses as low as 120~GeV, the resonance
contribution is quite noticeable and, because of the strong dependence on
$m_H$ of the $H\to WW$ branching ratio, this resonance contribution is
strongly dependent on the Higgs mass. When trying to show results for 
the $WW$ continuum only, we therefore impose the additional requirement
\beq
\label{eq:offres}
m_{WW} = \sqrt{(p_e+p_\mu+p_{\nu_e}+p_{\nu_\mu})^2}\, > \, m_H+10\;{\rm GeV}\;,
\eeq
i.e.\ the four-lepton invariant mass must be above the Higgs resonance.
The resulting cross section is representative of the continuum above 
any light Higgs boson resonance ($m_H$ below the $W$-pair threshold). 

\begin{figure}[!thb] 
\centerline{ 
\epsfig{figure=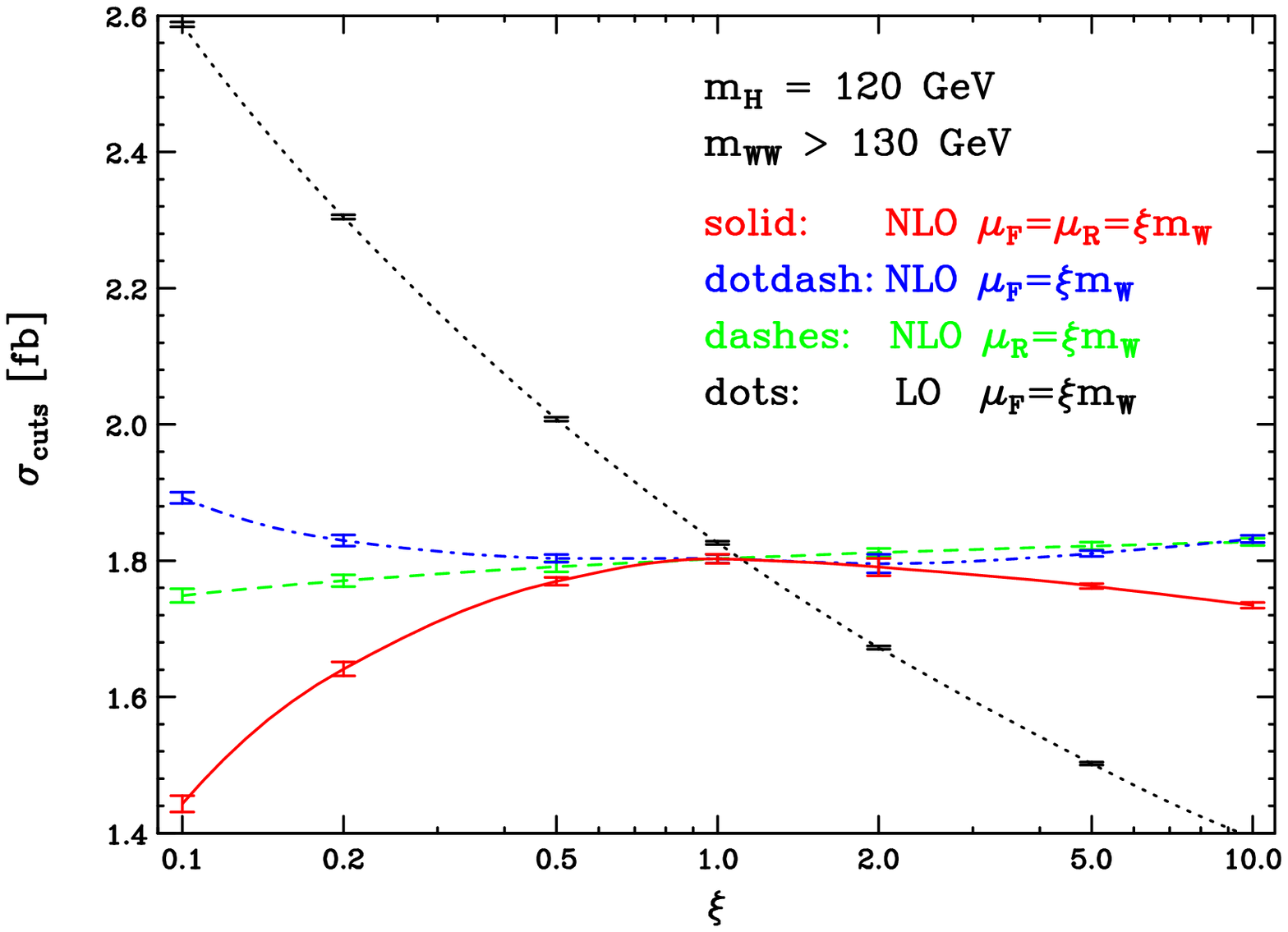,width=0.8\textwidth,clip=}
} 
\ccaption{} 
{\label{fig:scale_dep_offres} 
Scale dependence of the total $jj \, \wwdecay$ cross
section at leading  and next-to-leading order within the cuts of
Eqs.~(\ref{eq:cuts1})--(\ref{eq:offres}) for $pp$ collisions at 
the LHC. The contribution from the Higgs resonance (taken as
$m_H=120$~GeV) is excluded. 
The factorization scale $\mu_F$ and/or the renormalization scale $\mu_R$
are taken as multiples of the $W$ mass, $\xi\, m_W$, and
$\xi$ is varied in the range $0.1 < \xi < 10$. The NLO curves are for
$\mu_F=\mu_R=\xi m_W$ (solid red line), 
$\mu_F=m_W$ and $\mu_R=\xi\, m_W$ (dashed green line) and 
$\mu_R=m_W$ and $\mu_F$ variable (dot-dashed blue line).
The dotted black curve shows the dependence of the LO cross
section on the factorization scale. At this order, there is no dependence on
$\alpha_s(\mu_R)$.
}
\end{figure}

The scale dependence of the total continuum cross section, for a Higgs 
boson mass of $m_H=120$~GeV, is shown in Fig.~\ref{fig:scale_dep_offres}.
\begin{figure}[thb] 
\centerline{ 
\epsfig{figure=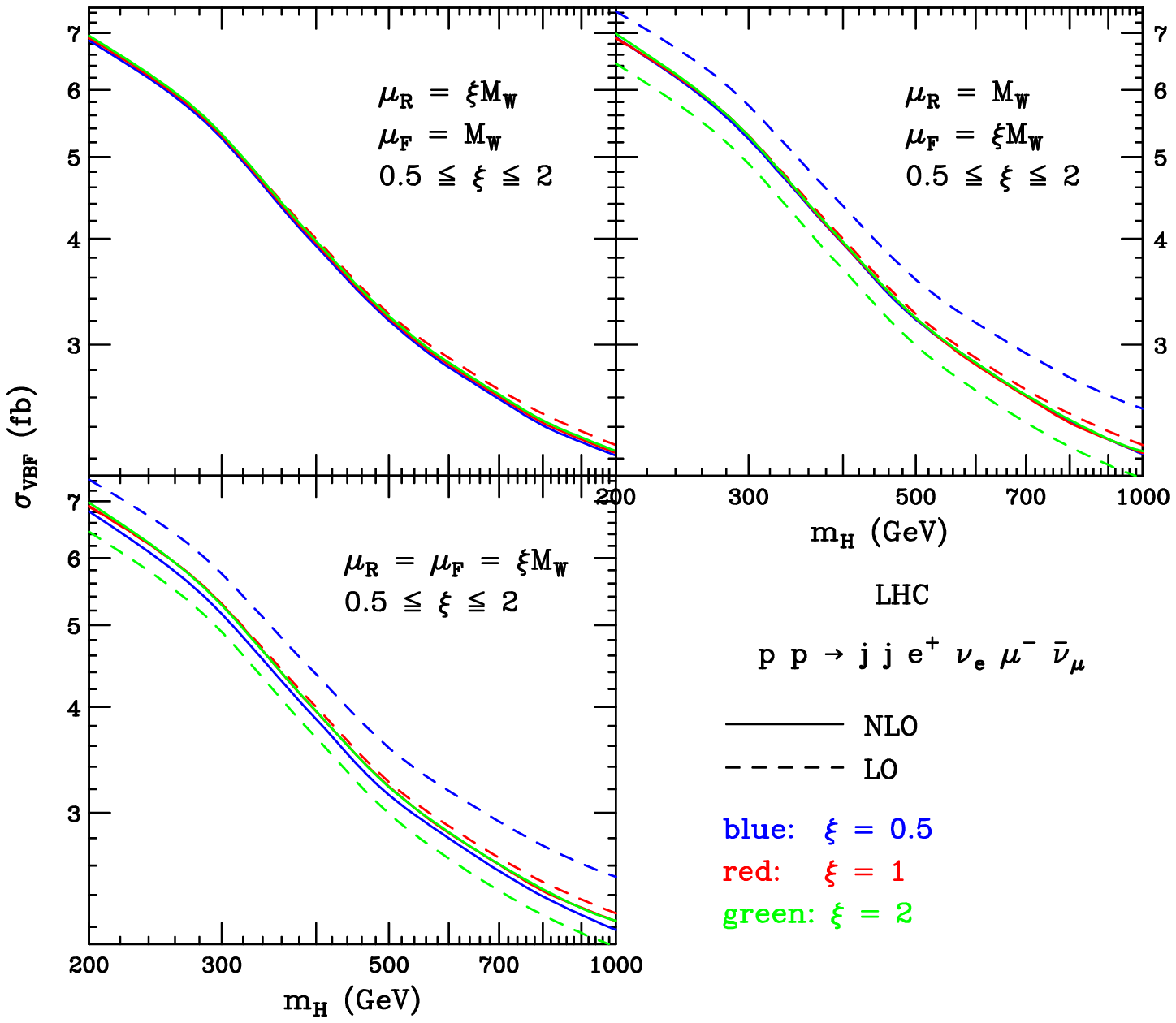,width=0.9\textwidth,clip=}%
} 
\ccaption{} 
{\label{fig:scale_dep_mH} 
Higgs mass dependence of the total $pp\,\to\, jj\, \wwdecay$ cross
section at LO and NLO within the cuts of
Eqs.~(\ref{eq:cuts1})--(\ref{eq:cuts6}). Results are shown for
renormalization and factorization scales $\mu = 0.5\;m_W$, $m_W$ and
$2\,m_W$.
}
\end{figure} 
This figure shows the 
scale dependence of the LO and NLO cross sections,
for renormalization and factorization scales, $\mu_R$ and
$\mu_F$, which are tied to the $W$ mass
\bq
\label{eq:scale.mV}
\mu_R = \xi_R\,m_W\;,\qquad\qquad \mu_F = \xi_F\,m_W\; .
\eq
The LO cross section only depends on the factorization scale. 
At NLO we show
three cases: (a)~$\xi_F=\xi_R=\xi$ (solid red line); (b)~variation of
the factorization scale only, $\xi_F=\xi$,
$\xi_R=1$ (dot-dashed blue line); and (c)~variation of the
renormalization scale only $\xi_R=\xi$, $\xi_F=1$ 
(dashed green line). The NLO cross sections are quite insensitive to
scale variations: 
allowing a factor 2 variation in either directions, i.e.\ considering the range
$0.5<\xi <2$, the NLO cross section changes by less than 2\% in all
cases. Compared to this small variation, the factorization scale
dependence of the LO cross section is quite sizable, amounting to a 
$\pm 10\%$ shift for $0.5<\xi <2$. Note that for $\mu_F=m_W$ the
LO cross section is only very slightly larger than the more stable NLO
result, yielding a \Kfac\  $K=\sigma_{NLO}/\sigma_{LO}=0.98$, i.e.\
$\mu_F=m_W$ is an excellent choice 
for a LO estimate of the total continuum cross section.

Also for larger Higgs boson masses, $m_H \gsim 2m_W$, the reduction of the
scale dependence at NLO is comparable to the light Higgs case. However,
since the resonance contribution can no longer be trivially separated from the 
$WW$ continuum, we now show, in Fig.~\ref{fig:scale_dep_mH}, the total
cross section within the cuts of Eqs.~(\ref{eq:cuts1})--(\ref{eq:cuts6})
as a function of $m_H$ and for different scale choices, $\mu = \xi m_W$
with $\xi=0.5,\;1$ and 2. At NLO, the scale dependence is hardly visible
while at LO one again finds a sizable factorization scale dependence.
 
The small scale dependence which is observed for the total cross section
at NLO is also found for infrared-safe distributions. Typically, scale
variations between $0.5\,m_W$ and $2\,m_W$ change distributions by about
2\%, with somewhat larger variations, up to 6\%, sometimes occurring in
the tails of the distributions shown below.   

The \Kfac\ close to unity, which was found for the total cross section,
no longer persists for distributions. We demonstrate this effect by
showing a few experimentally relevant distributions together with the
dynamic \Kfac\ which is defined as
\bq
\label{eq:kfactor}
K(x) = \frac{d\sigma_{NLO}/dx}{d\sigma_{LO}/dx}\;.
\eq
In the following the Higgs boson mass is taken as $m_H=120$~GeV and we 
show cross sections for the continuum above $m_{WW}=130$~GeV and within 
the cuts of Eqs.~(\ref{eq:cuts1})--(\ref{eq:offres}). All panels are for the
scale choice $\mu_F=\mu_R=m_W$.

\begin{figure}[thb] 
\centerline{ 
\epsfig{figure=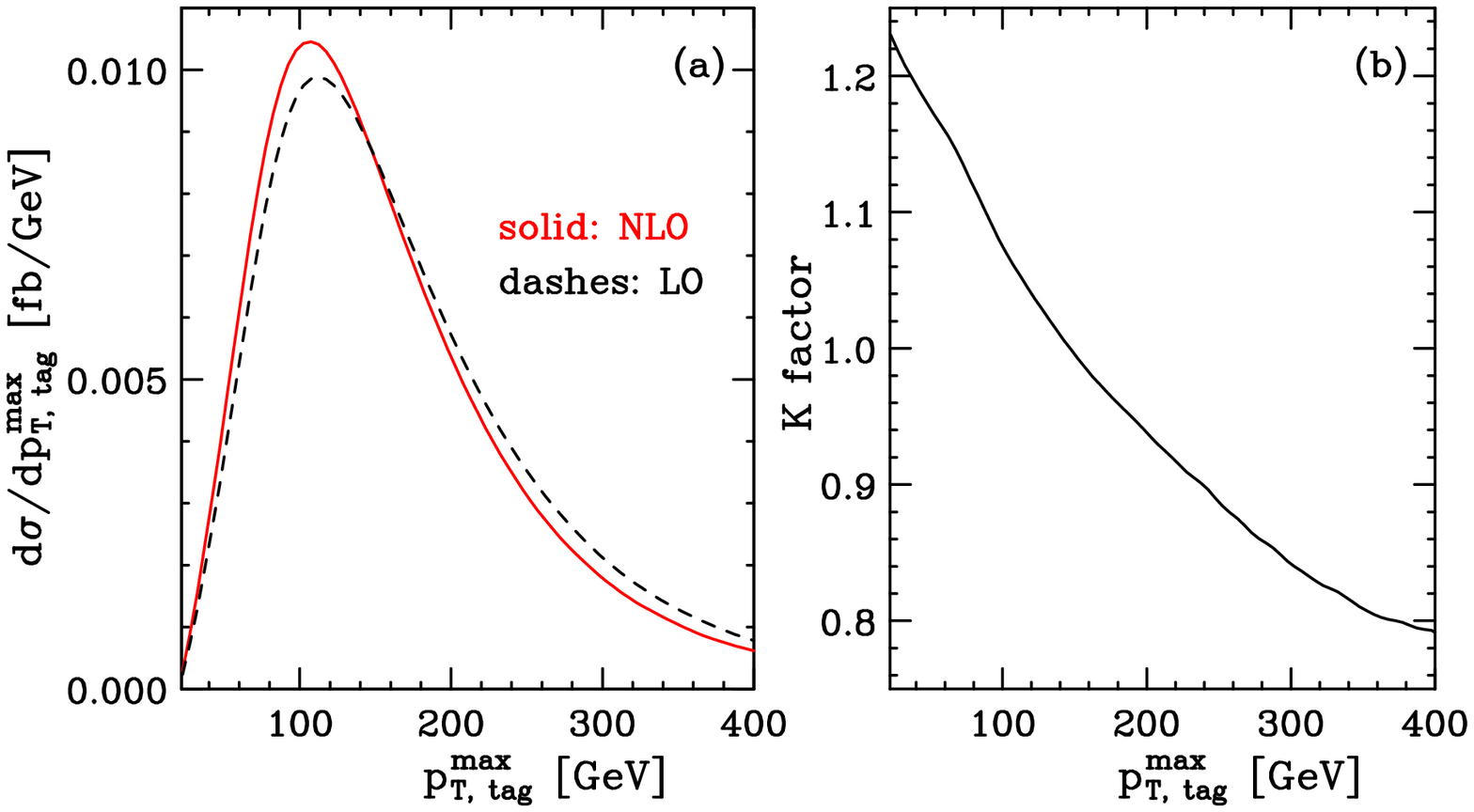,width=0.9\textwidth,clip=} 
} 
\ccaption{} 
{\label{fig:pt_max_tagj}
Transverse-momentum distribution of the highest-$p_T$ 
tagging jet in EW $\wwjj$ production at the LHC. In panel~(a) the NLO result
(solid red line) and the LO curve (dashed black line) are shown. 
Their ratio, the \Kfac\  as defined in Eq.~(\ref{eq:kfactor}), is shown in 
panel~(b).
}
\end{figure}

\begin{figure}[htb] 
\centerline{ 
\epsfig{figure=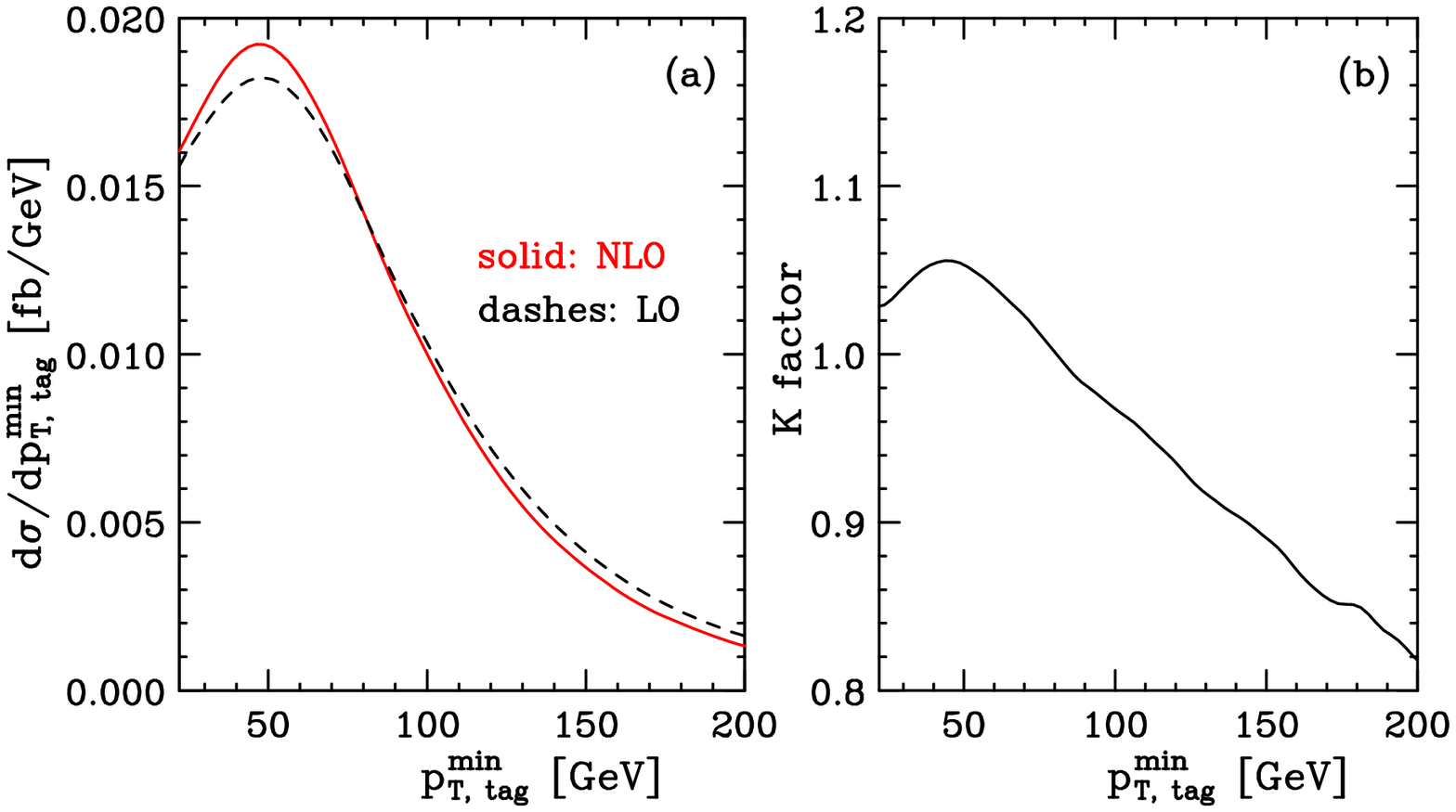,width=0.9\textwidth,clip=} 
} 
\ccaption{} 
{\label{fig:pt_min_tagj} 
Same as Fig.~\ref{fig:pt_max_tagj} but for the smaller of the two tagging-jet
transverse momenta. 
}
\end{figure} 

\begin{figure}[!thb] 
\centerline{ 
\epsfig{figure=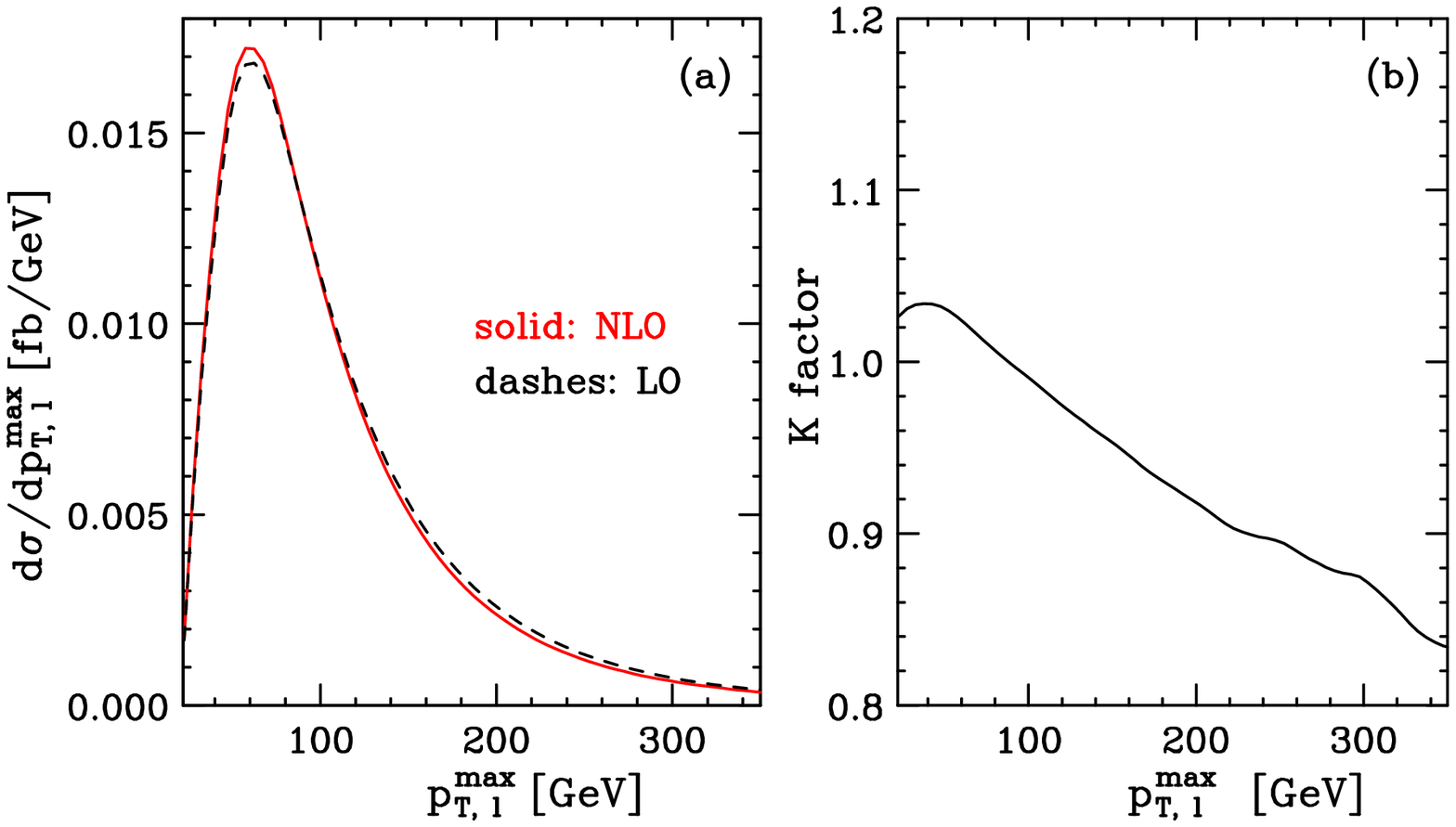,width=0.88\textwidth,clip=} 
} 
\ccaption{} 
{\label{fig:pt_L_max}
Transverse-momentum distributions of the hardest final-state lepton in 
EW $\wwjj$ production at the LHC. In panel~(a) the NLO result
(solid red line) and the LO curve (dashed black line) are shown. 
Their ratio, the \Kfac\  as defined in Eq.~(\ref{eq:kfactor}), is shown in 
panel~(b).
}
\end{figure}

A fairly strong shape change in going from LO to NLO is found for the 
tagging-jet transverse-momentum distributions. This is shown in 
Figs.~\ref{fig:pt_max_tagj} and~\ref{fig:pt_min_tagj} where the 
larger and the smaller of the two tagging-jet transverse momenta are shown
at LO (dashed black curves) and at NLO QCD (solid red lines), together 
with their ratio, the \Kfac\  of Eq.~(\ref{eq:kfactor}). In particular 
the former, $d\sigma/dp_{T,{\rm tag}}^{{\rm max}}$, shows a clear shift to smaller 
transverse momenta at NLO, which corresponds to a \Kfac\  varying between
1.2 and 0.8 as $p_{T,{\rm tag}}^{{\rm max}}$ increases from 20~GeV to 400~GeV. 
The effect for $d\sigma/dp_{T,{\rm tag}}^{\rm {min}}$, in
Fig.~\ref{fig:pt_min_tagj}, is slightly smaller, but still pronounced.
The change in the jet transverse-momentum distribution also feeds into the 
shape of the lepton transverse-momentum distributions. In 
Fig.~\ref{fig:pt_L_max} we depict the transverse momentum for the hardest 
of the two charged leptons. Again small transverse momenta are enhanced at 
NLO, leading to a \Kfac\ between 1.04 and 0.84.

\begin{figure}[!thb] 
\centerline{ 
\epsfig{figure=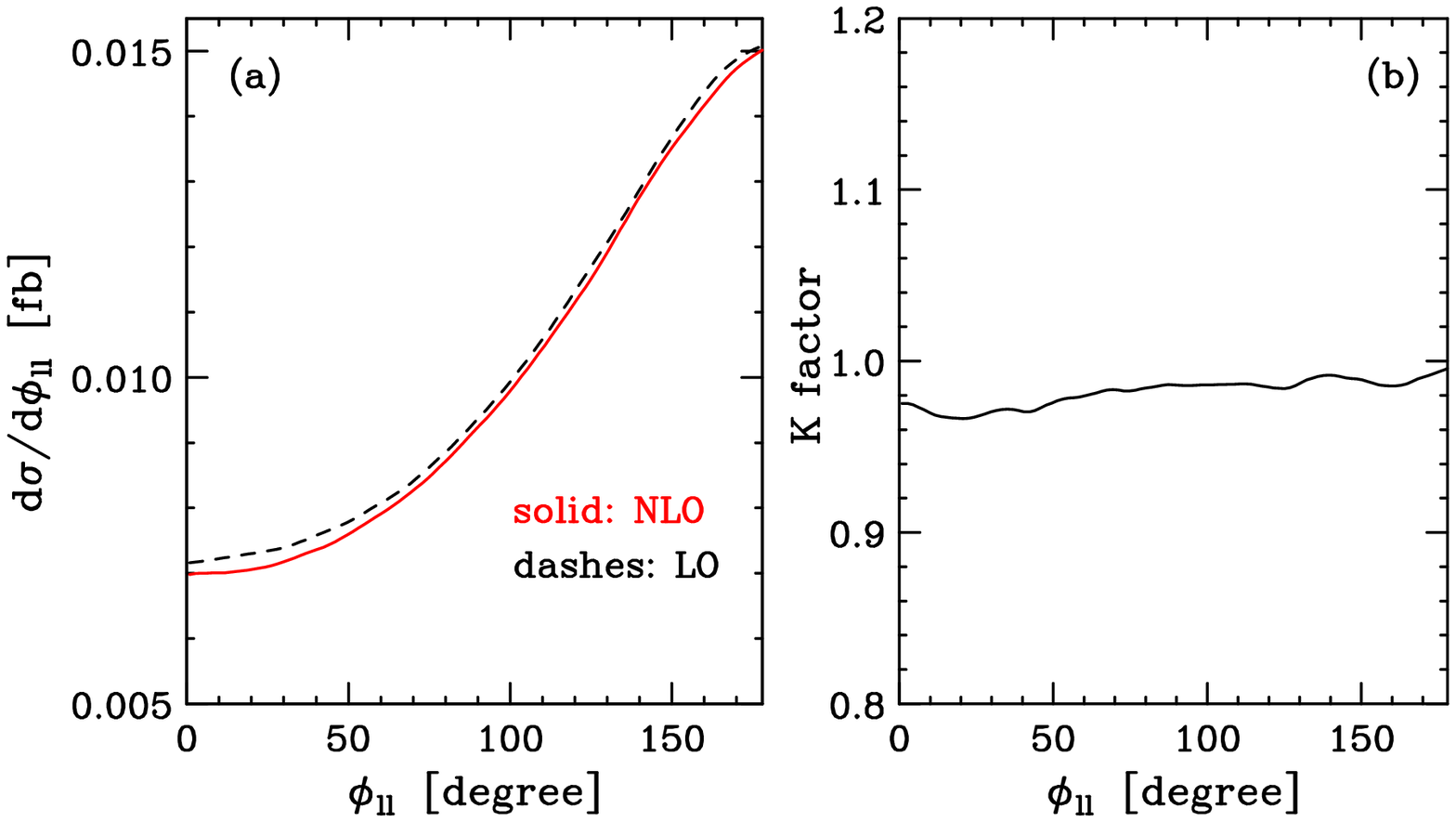,width=0.9\textwidth,clip=}
} 
\ccaption{} 
{\label{fig:phill} 
Azimuthal angle separation between the two charged leptons for continuum
$jj\wwdecay$ events at the LHC. Curves are as in Fig.~\ref{fig:pt_max_tagj}.
}
\end{figure}

In contrast to the transverse momentum distributions, angular distributions 
of the leptons are hardly affected by the NLO corrections. As an example, we 
show the azimuthal angle between the two charged leptons in 
Fig.~\ref{fig:phill}. The \Kfac\  is almost constant and equal to
$0.98$. The typically large angle between the charged leptons  
is important for the reduction of $\wwjj$ continuum events in the search for
$H\to WW\to l^+l^-\sla{p}_T$ decays~\cite{wbfhtoww}.

\begin{figure}[thb] 
\centerline{ 
\epsfig{figure=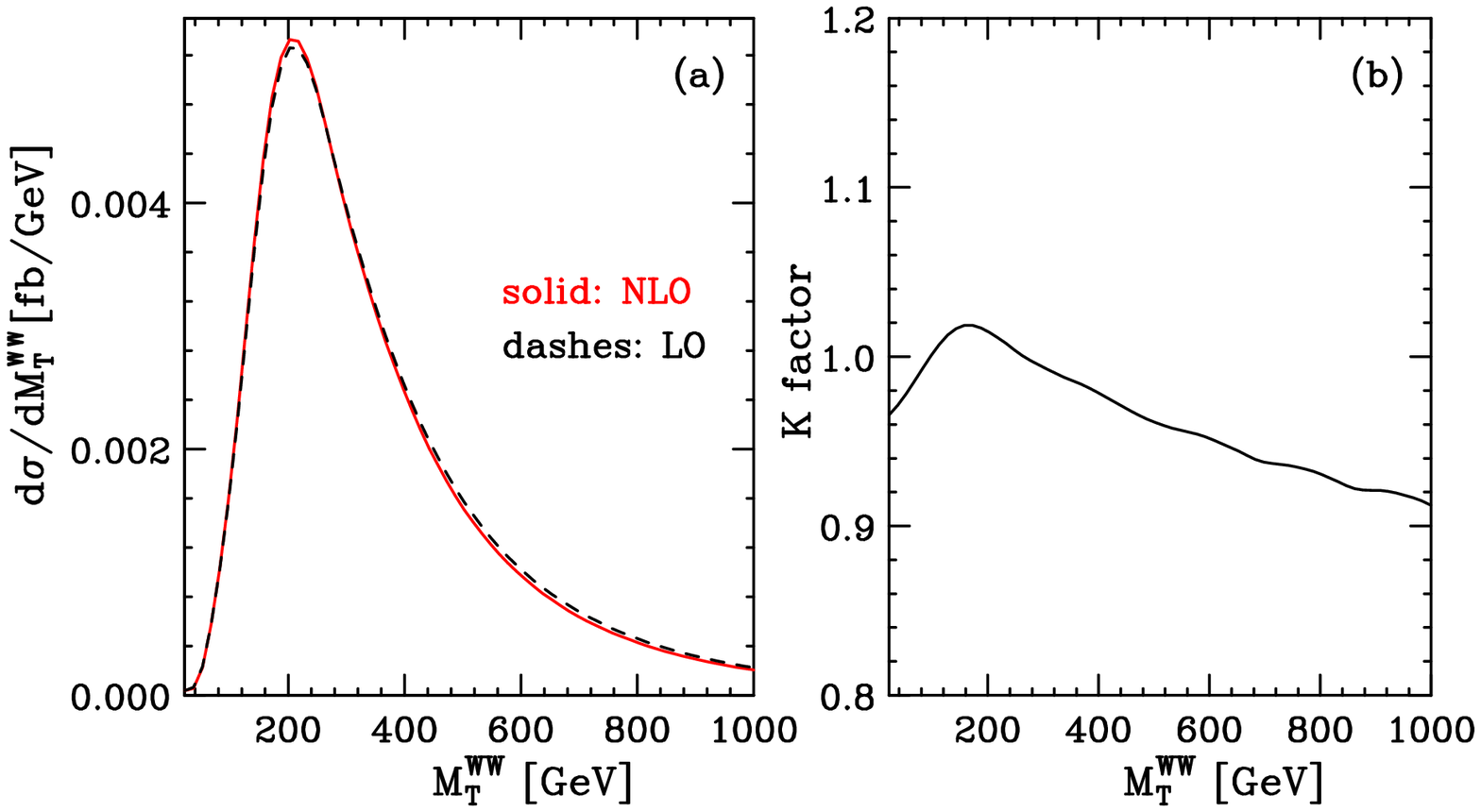,width=0.9\textwidth,clip=} 
} 
\ccaption{} 
{\label{fig:mt_ww} 
Transverse mass distribution for the $\wwdecay$ system in
$\wwjj$ events at the LHC. Curves are as in Fig.~\ref{fig:pt_max_tagj}.
The definition of $M_T^{WW}$ is given in Eq.~(\protect{\ref{eq:M_T^WW}}).
}
\end{figure}

Another distribution which is important for the Higgs search at the LHC 
is the transverse mass of the $l^+l^-\nu\bar\nu$ system, which is defined 
as~\cite{wbfhtoww}
\beq
\label{eq:M_T^WW}
M_T^{WW} = 
\sqrt{({\sla{E}_T}+E_{T,ll})^2 - ({\bf p}_{T,ll}+{\sla{\bf p}}_T)^2}\,,
\eeq 
where the transverse energies are given by
\beqn
E_{T,ll} &=& \sqrt{{\bf p}_{T,ll}^2 + m_{ll}^2}\,, \nonumber \\
\sla{E}_T &=& \sqrt{{\sla{\bf p}}_T^2 + m_{\nu\nu}^2} \approx
\sqrt{{\sla{\bf p}}_T^2 + m_{ll}^2} \,.
\eeqn 
While the invariant mass of the $W^+W^-$ pair cannot be reconstructed, due to
the presence of neutrinos, $M_T^{WW}$ is fully accessible.  The effect of NLO
QCD corrections on $M_T^{WW}$ is again modest, as can be seen in
Fig.~\ref{fig:mt_ww}.

\section{Conclusions}
\label{sec:summary}
Vector-boson fusion at the LHC represents a class of electroweak processes 
which are under excellent control perturbatively. This has been known for
some time for the most interesting process in this class: 
Higgs boson production via VBF has a modest \Kfac\ of about 1.05
for the inclusive production cross section~\cite{Han:1992hr} and this result
also holds when applying realistic acceptance
cuts~\cite{Figy:2003nv}. Similar results were also found for $Wjj$ and
$Zjj$ production via VBF~\cite{Oleari:2003tc}.

In the present paper, we have extended these calculations to the electroweak
process $pp\,\to\,\wwdecay \,jj$ at NLO in QCD, when the
final-state particles are in a kinematic configuration typical of VBF
events. This corresponds to leptonic final states in the vector-boson
scattering processes $VV\to W^+W^-$ ($V$ is a $\gamma$ or a $Z$) and
$W^+W^-\to W^+W^-$, but with full 
NLO QCD simulation of the associated tagging jets. The calculation has
been implemented in the form of a fully-flexible parton level Monte
Carlo program and, thus, allows to implement completely general
experimental cuts. The size of the QCD corrections is similar to those
found for $Hjj$ and $Vjj$ production in VBF, and corresponds to a shift of
a few percent in typical integrated cross sections expected for VBF
cuts. Some distributions, however, are affected somewhat more strongly, with
dynamical {\Kfac s}  ranging between 0.8 and 1.2, in particular for
transverse-momentum distributions. At least as important is the 
stability of the NLO result: the residual scale dependence is at the 2\%
level for cross sections integrated within VBF cuts.  

The numerical code is quite fast, reaching permille level statistics on
distributions within 5 days of running on a standard 3~GHz PC. A
$1\permil$ error on integrated cross sections is reached in about 1
day.  This high speed has been obtained by avoiding the recalculation of
recurring subamplitudes in 
different sub-processes contributing at a given phase-space point. A key
ingredient is a modular structure of the numerical amplitude calculation
which separates the 
weak-boson scattering sub-amplitudes into leptonic tensors, which can be
changed without altering the validity of the QCD corrections. Such
changes could reflect the inclusion of anomalous three- or
four-vector-boson couplings or of any other new physics in weak-boson
scattering. We leave such generalizations for the future.


\section*{Acknowledgments}

This research was supported in part by the Deutsche
Forschungsgemeinschaft in the Sonderforschungsbereich/Transregio 
SFB/TR-9 ``Computational Particle Physics''.

\end{document}